\documentclass[reprint]{revtex4-2}
\usepackage[utf8]{inputenc}
\usepackage[T1]{fontenc}
\usepackage{array,amsmath,amssymb,siunitx,physics}
\usepackage{xcolor,graphicx}
\usepackage{ascmac,enumitem,autobreak}

\begin{document}
\title{Predicted diversity in water content of 
terrestrial exoplanets orbiting M dwarfs}

\author{Tadahiro Kimura}
\email{t.kimura@eps.s.u-tokyo.ac.jp}
\affiliation{Department of Earth and Planetary Science, Graduate School of Science, The University of Tokyo, 7-3-1 Hongo, Bunkyo-ku, Tokyo 113-0033, Japan}
\author{Masahiro Ikoma}
\email{ikoma.masahiro@gmail.com}
\affiliation{Division of Science, National Astronomical Observatory of Japan, 2-21-1 Osawa, Mitaka, Tokyo 181-8558, Japan}
\affiliation{Department of Earth and Planetary Science, Graduate School of Science, The University of Tokyo, 7-3-1 Hongo, Bunkyo-ku, Tokyo 113-0033, Japan}
\affiliation{Department of Astronomical Science, The Graduate University for Advanced Studies (SOKENDAI), 2-21-1 Osawa, Mitaka, Tokyo 181-8558, Japan}

\maketitle

\textbf{
Exoplanet surveys around M dwarfs have detected a growing number of exoplanets with Earth-like insolation. It is expected that some of those planets are rocky planets with the potential for temperate climates favourable to surface liquid water. However, various models predict that terrestrial planets orbiting in the classical habitable zone around M dwarfs have no water or too much water, suggesting that habitable planets around M dwarfs might be rare. Here we present the results of an updated planetary population synthesis model, which includes the effects of water enrichment in the primordial atmosphere, caused by the oxidation of atmospheric hydrogen by rocky materials from incoming planetesimals and from the magma ocean. We find that this water production in the primordial atmosphere is found to significantly impact the occurrence of terrestrial rocky aqua planets, yielding ones with diverse water content. We estimate that 5--10\% of the planets with a size $<1.3 R_\oplus$ orbiting early-to-mid M dwarfs have 
appropriate amounts of seawater for habitability.
Such an occurrence rate would be high enough to detect potentially habitable planets by ongoing and near-future M-dwarf planet survey missions. 
}
\section*{Introduction}
The Earth's temperate climate has been maintained through the geochemical carbon cycle in which weathering plays a key role \cite{Walker+1981}. 
On the present-day Earth, weathering takes place efficiently on land. 
Lands exist on the Earth because the planet has a moderate amount of seawater accounting for 0.023\% of the planet's total mass. 
On planets having tens of times more seawater than the Earth, weathering could not work efficiently enough to maintain temperate climates (e.g., refs.\cite{Abbot+2012,Alibert2014,Nakayama+2019}).  
Regarding the origin of seawater, a widespread idea is that the Earth's seawater was brought by water-laden or icy planetesimals (e.g., ref.\cite{Genda2016}).
Based on this idea, one would naturally predict a bimodal distribution of planetary water contents, since initial planetesimals are distinctly different in water content between the regions interior and exterior to the snowline in a protoplanetary disc, which was demonstrated by refs.\cite{Tian+Ida2015,Miguel+2019}, as described above.

Alternatively, water can be secondarily produced in a primordial atmosphere of nebular origin through reaction of atmospheric hydrogen with oxidising minerals from the magma ocean, which is formed because of the atmospheric blanketing effect\cite{Ikoma+Genda2006}, thereby enriching the primordial atmosphere with water.
By assuming effective water production, we recently showed that nearly-Earth-mass planets can acquire sufficient amounts of water for their atmospheric vapour to survive in harsh UV environments around pre-main-sequence M stars~\cite{Kimura+Ikoma2020}.
The results suggest that including this water production process significantly affects the predicted water amount distribution of exoplanets in the habitable zone around M dwarfs.

Our planetary population synthesis model, which follows the evolution of planets' masses, radii, and orbits by combining empirical laws for several components of the planet formation process based on the planetesimal accretion hypothesis,
is similar to those of refs.~\cite{Ida+2013,Ida+2018,Miguel+2019} but includes the movement of snowline location due to the thermal evolution of the protoplanetary disc, the accumulation of atmospheres of nebular origin (i.e., the primordial atmospheres) also before the core mass becomes critical, and the effects of water production in the primordial atmospheres\cite{Kimura+Ikoma2020}. 
Furthermore we have updated the treatment of some of the processes involved in planet formation/evolution according to improved understandings (See Method). 
In this study, we have performed Monte Carlo calculations for $1 \times 10^9$~yr with 10000 different initial conditions for a given set of the input parameters (See Method), and investigated the frequency distribution of water content among the synthesised planets. 

\section*{Results}
\subsection*{Planetary mass and semi-major axis distribution}
\begin{figure*}
    \centering
    \includegraphics{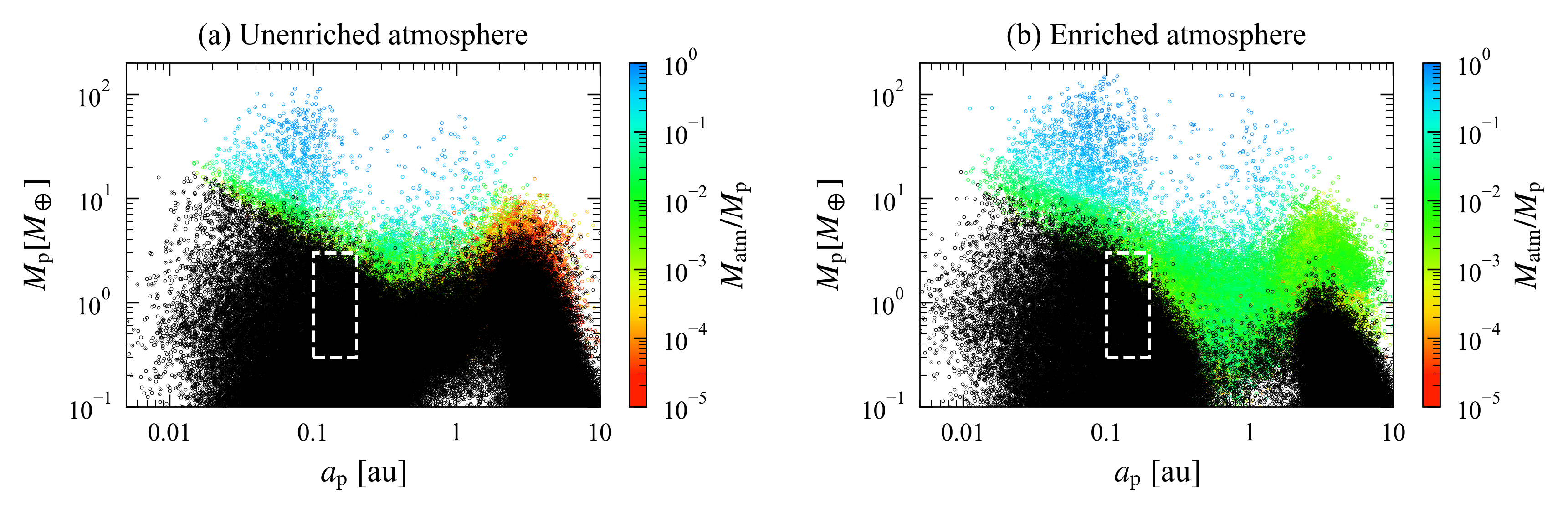}    
    \caption{Effects of water enrichment in the primordial atmospheres on atmospheric accumulation and planetary growth. The two panels show the
    planet mass ($M_\mathrm{p}$) vs. semi-major axis ($a_\mathrm{p}$) distributions of the synthesised planets orbiting $0.3M_\odot$ stars at 1~Gyr after the beginning of calculations for the water mass fraction in the primordial atmosphere of (a) $X_{\text{H}_2\text{O}}=0.0$ and (b) $X_{\text{H}_2\text{O}}=0.8$, respectively.
    The symbols are colour-coded according to the
    atmospheric mass ($M_\text{atm}$) relative to the planet's total mass ($M_\text{p}$).
    Note that planets without any atmosphere are shown in black.
    The dashed box in each panel shows the region of nearly Earth-mass planets in the present-day habitable zone (HZ-NEMPs; see text for the definition).
    }
    \label{fig:Ma_M03}
\end{figure*} 
Firstly our planetary population synthesis calculations demonstrate that water enrichment in the primordial atmosphere has 
a great effect on atmospheric accumulation 
of low-mass planets such as sub-Earths and super-Earth (see Fig.~\ref{fig:Ma_M03}).
The overall distributions of planetary masses, $M_\text{p}$, and semi-major axes, $a$, for unenriched (panel (a)) and enriched (panel (b)) atmospheres are similar to each other. However, it turns out that sub/super-Earths with enriched atmospheres consequently have more massive atmospheres in relatively cool regions (see green symbols in panel (b)). This is because H$_2$O enrichment leads to increasing the atmospheric mean molecular weight and the effective heat capacity through condensation and chemical reactions, making the atmosphere denser \cite{Kimura+Ikoma2020}. 

\subsection*{Water content distribution}
\begin{figure}
    \centering
    \includegraphics{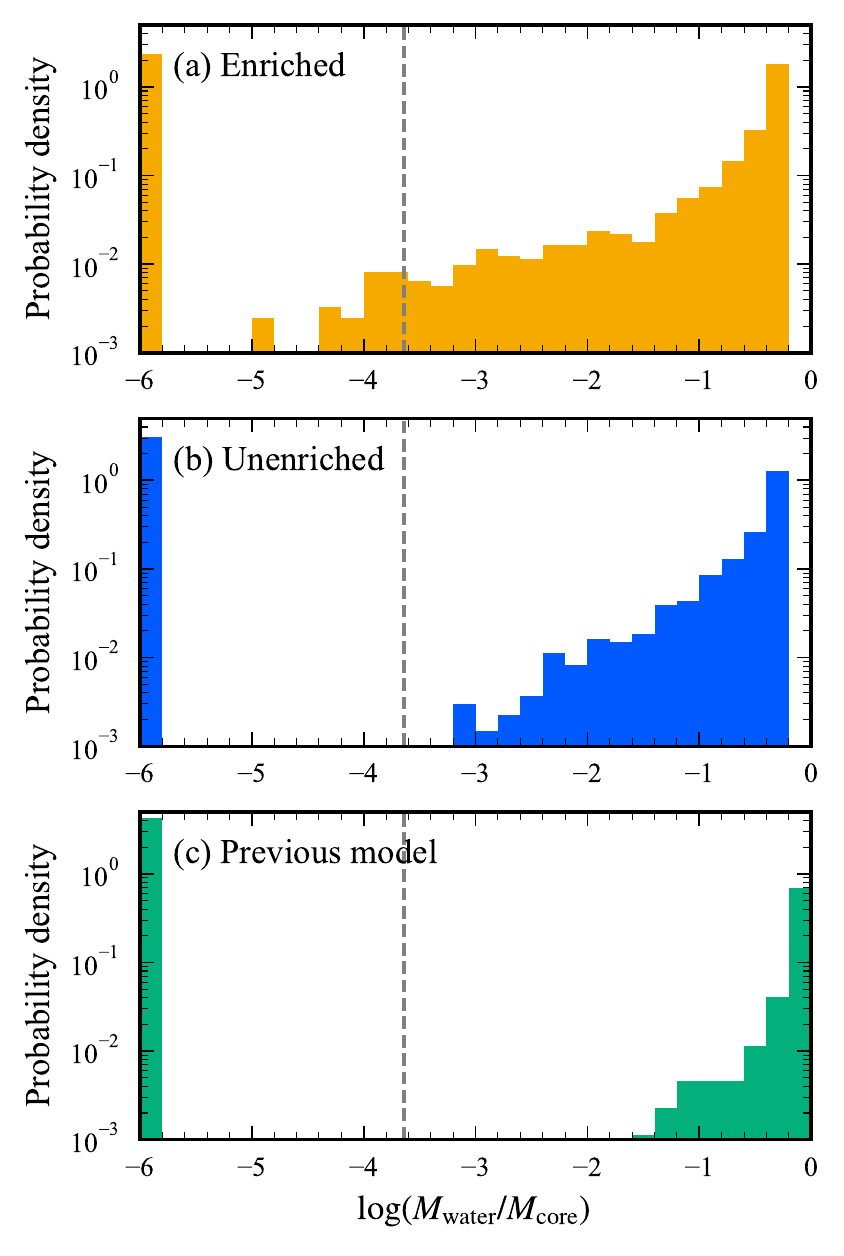}
    \caption{Probability density distribution of the water mass fraction in the synthesized planets with mass of $0.3$--$3M_\oplus$ ("nearly Earth-mass planets") orbiting at between 0.1 and 0.2~au ("habitable zone") around M dwarfs of $0.3M_\odot$ at the age of 1~Gyr. 
    The panels show the results for the cases with (a) enriched atmospheres ($X_{\rm H_2O}=0.8$), (b) unenriched atmospheres ($X_{\rm H_2O}=0.0$), and (c) the same assumptions and settings as ref.~\cite{Tian+Ida2015}.
    Planets without any water are shown in the leftmost bar in each panel for drawing purpose.
    The vertical dashed line indicates the ocean mass fraction for the present-day Earth ($\num{2.3e-4}$).}
    \label{fig:hist_M03}
\end{figure}
\begin{figure}
    \centering
    \includegraphics{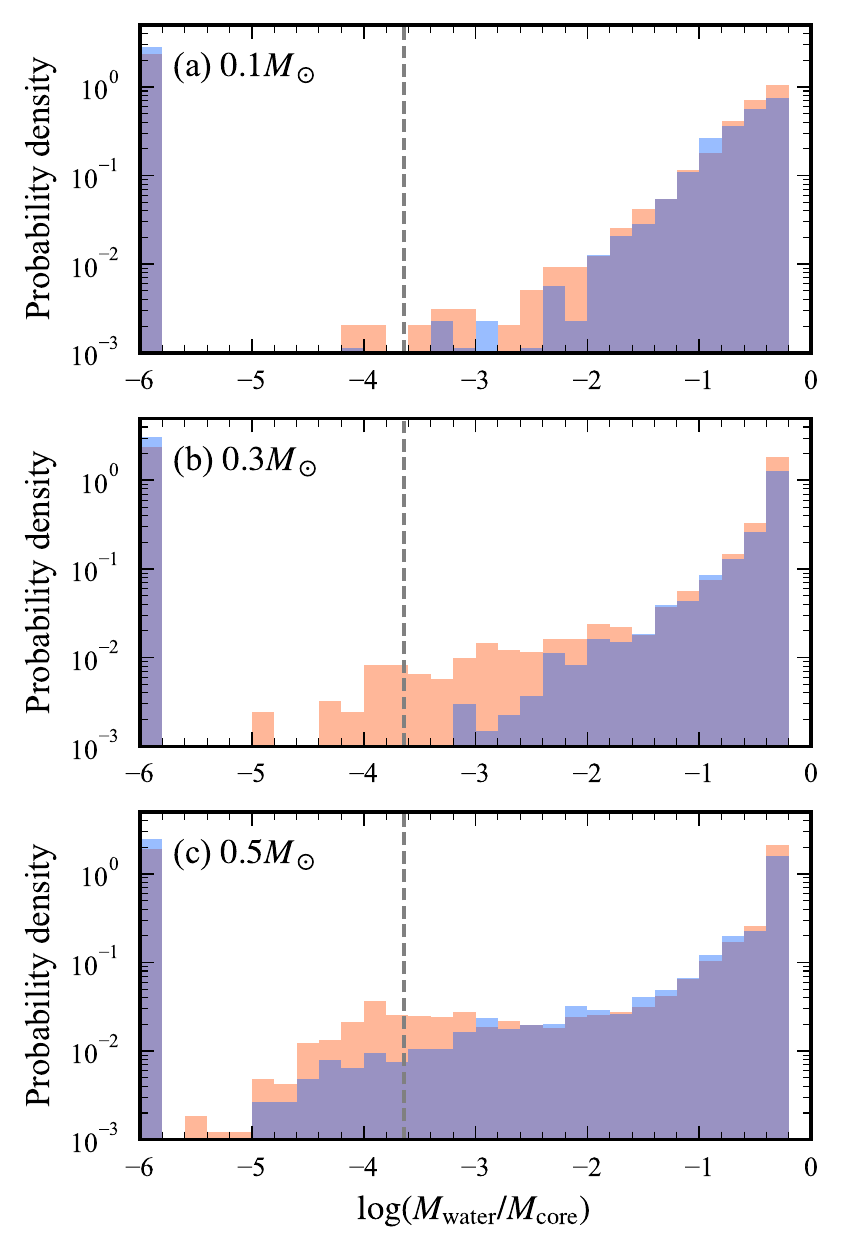}
    \caption{Same as Fig.~\ref{fig:hist_M03}~(a) and (b), but for different stellar masses, (a) 0.1$M_\odot$, (b) 0.3$M_\odot$, and (c) 0.5$M_\odot$.
    The orange and blue bars show the results for enriched and unenriched atmospheres (water mass fraction in the primordial atmosphere being $X_{\text{H}_2\text{O}}=0.8$ and 0.0), respectively.}
    \label{fig:hist_Mstar}
\end{figure}
Such an increase of low-mass planets with relatively massive H$_2$O-enriched atmospheres greatly affects the occurrence of aqua planets in the habitable zones around M dwarfs. 
Figure~\ref{fig:hist_M03} shows the probability density distributions of the water mass fraction in planets with mass of 0.3--3~$M_\oplus$ in the \textit{present-day} habitable zone (HZ) around 0.3~$M_\odot$ M dwarfs for the cases with enriched ($X_{\text{H}_2\text{O}}=0.8$; panel (a)) and unenriched ($X_{\text{H}_2\text{O}}=0.0$; panel (b)) atmospheres (See \textit{Method} for the definition of the density probability). 
Such planets are hereafter called the habitable-zone nearly-Earth-mass planets (HZ-NEMPs). 
Note that the present-day habitable zone is defined as the habitable zone around the star with age of 5~Gyr.
For comparison, the result obtained under the same assumptions and settings as ref.~\cite{Tian+Ida2015} is also shown in panel (c).
Firstly, compared to the previous model~\cite{Tian+Ida2015} (panel (c)), 
our planetary population synthesis models produce HZ-NEMPs with a much wider range of water contents even for $X_{\text{H}_2\text{O}}=0.0$ (in particular, those with water mass fraction of $\lesssim$10\%). 
This is due to the shift in location of the snowline with the protoplanetary disc's thermal evolution, which is not included in the previous model.
Initially, when viscous heating dominates, the snowline is located at $\sim 1$~au.
As viscous heating diminishes and, then, stellar irradiation dominates, the snowline migrates inward, reaching $\sim 0.2$~au, corresponding to the outer edge of the prersent-day HZ, on a timescale of Myrs (see Supplementary Figure 1 and Section 1.1 in the Supplementary Information for details).
Here, our model assumes that 
once the snowline passes through inward, water vapour immediately condenses onto rocky planetesimals, making planetesimals with an ice-rock ratio of unity.
Thus, planets originally formed in this region can acquire small amounts of ice, depending on the timescale of planetary mass growth, orbital migration, and the snowline migration.
Effects of this assumption are discussed in Section 1.2 in the Supplementary Information, with Supplement Fig.2.

The water enrichment in the primordial atmosphere, which is of special interest in this study, is found to bring about further increase in water contents of HZ-NEMPs (see panel~(a) of Fig.~\ref{fig:hist_M03}). 
Because of the enrichment, even without capturing icy planetesimals, the rocky planets obtain water from their primordial atmosphere. 
As demonstrated in Fig.~\ref{fig:Ma_M03} (b), the water enrichment enhances the accumulation of the primordial atmospheres. Consequently, the captured H$_2$/He and secondarily produced H$_2$O are large in amount enough to survive the subsequent atmospheric photo-evaporation process. This mechanism forms rocky planets with water mass fraction of $< 1~\%$ inside the snowline.
In contrast to the previous models\cite{Tian+Ida2015,Miguel+2019} which predict the absence of HZ-NEMPs with the Earth-like water content ($2.3 \times 10^{-4}$), our model with enriched primordial atmospheres predicts that a significant number of such HZ-NEMPs are formed. 
It is noted that the abundance of HZ-NEMPs with Earth-like water contents is significantly affected by the protoplanetary disc conditions, especially the disc lifetime.
Most of the HZ-NEMPs with small amounts ($\lesssim$ 1~wt.\%) of water are found to form in discs with lifetime of $\lesssim$ 3~Myr.
In longer-lived discs, the outer icy planets significantly migrate inward, pushing the inner rocky planets closer to the disc inner edge.
As a result, only ice-dominant planets exist in HZ.

\subsection*{Dependence on central stellar mass}

Similarly to Fig.~\ref{fig:hist_M03}, Fig.~\ref{fig:hist_Mstar} shows the probability density distribution for HZ-NEMPs orbiting stars of three different masses (0.1, 0.3 and 0.5$M_\odot$), around which the habitable zones are located at 0.03--0.07~au, 0.1--0.2~au and 0.2--0.4~au, respectively~\cite{Kopparapu+2014}, in their main-sequence phase.
As found in this figure, the larger the stellar mass, the higher the abundance of HZ-NEMPs with small water contents ($\lesssim$ 1~wt.\%).
This is due to two effects: Firstly, the region that corresponds to the HZ when the host star is on its main sequence is located at a larger orbital distance for larger stellar mass. 
In that region of a protoplanetary disc, therefore, protoplanets can grow larger around more massive stars;
consequently, the protoplanets can obtain more atmospheric gas, thereby producing more water.
Secondly, the less massive the host star, the higher the stellar XUV irradiation in the habitable zone during the pre-main sequence phase. Therefore, the planets in the HZ undergo severer loss of the atmosphere and are less likely to keep water.
In the cases with unenriched atmospheres, the distribution also differs depending on stellar mass (dashed bars).
The distributions for 0.1 and 0.3$M_\odot$ stars are similar to each other, whereas the planets around 0.5$M_\odot$ stars have a much wider range of the water contents, and the distribution is rather similar to that for enriched atmospheres.
This indicates that many of the HZ-NEMPs around 0.5$M_\odot$ stars have obtained relatively large amounts of icy planetesimals during their formation.
This is because the snowline comes closer to the habitable zone around more massive stars; for 0.5$M_\odot$ stars, the former comes inside the latter.

\section*{Discussion}
\begin{figure*}
    \centering
    \includegraphics{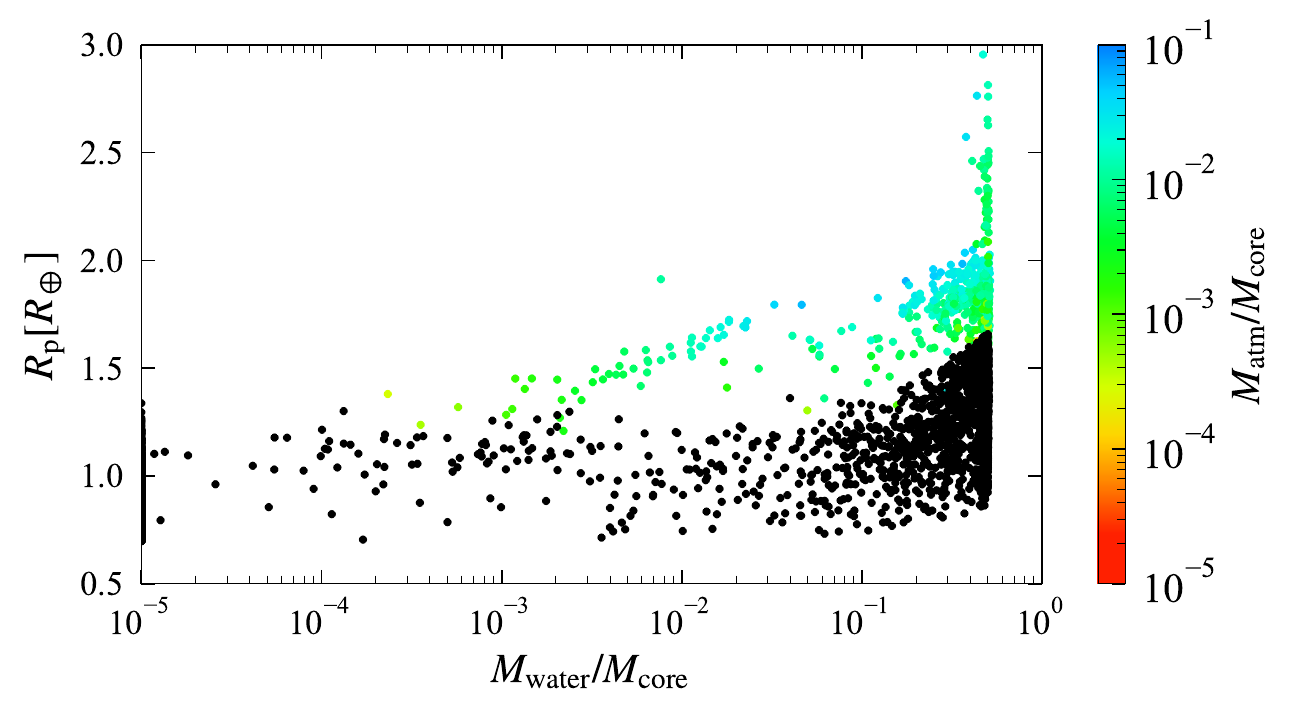}
    \caption{Planetary radius, $R_\text{p}$ vs. the water content of the ``nearly Earth-mass planets'' in the habitable zone (HZ-NEMPs) around $0.3M_\odot$ M dwarfs in the case of enriched atmospheres with the primordial-atmospheric water mass fraction $X_{\rm H_2O}=0.8$.
    The symbols are colour-coded according to the atmospheric mass relative to the planetary core mass.
    Atmosphere-free planets are shown by black points.
    Planets without any water are piled up at $M_\text{water}/M_\text{core}=10^{-5}$ for drawing purposes.
    }
    \label{fig:Rp_Fwater}
\end{figure*}
Our planetary population synthesis calculations have shown that the HZ-NEMPs orbiting M dwarfs can have diverse water contents by the effects of the disc's thermal evolution and the water enrichment in the primordial atmosphere.
Especially, the latter effect enables about 1\% of the HZ-NEMPs to have water amounts comparable to the Earth oceans, provided the atmospheric water mass fraction of 80~\% is achieved.
When water is produced through the chemical reaction between the atmospheric hydrogen and oxides in the magma ocean~\cite{Ikoma+Genda2006}, the resultant water mass fraction $X_{\text{H}_2\text{O}}$ depends on the kind of oxide; the equilibrium partial pressure ratio $P_{\text{H}_2\text{O}}/P_{\text{H}_2}$ is on the order of unity if the normal iron oxides such as w\"{u}stite are available from the magma.
The ratio $P_{\text{H}_2\text{O}}/P_{\text{H}_2}=1$ corresponds to about 80 wt.\% of H$_2$O in the H$_2$-He-H$_2$O atmosphere with the solar He/H ratio (=0.385~\cite{Lodders+2009}).
Thus, the assumed value of $X_{\text{H}_2\text{O}}=0.8$ is feasible if the entire atmospheric hydrogen is equilibrated with the oxides.
We should note that our results and conclusion hardly change for $X_{\text{H}_2\text{O}} \gtrsim 0.5$, because the atmospheric mass significantly increases in that range (see ref.~\cite{Kimura+Ikoma2020} for details).
We further discuss the implication of this water production process in the primordial atmosphere for the Earth in Section 2.3 in the Supplement Information.

Our planetary population synthesis model is based on the planetesimal accretion scenario, and does not consider pebbles.
The pebble accretion may affect our results and conclusion mainly in the following three ways: changes in the disc gas composition, in the oxidation state of the planetary rocky components, and in the planetary mass distribution.
Firstly, drifting pebbles can enhance the water content of disc gas inside the snowline through the sublimation of ice~\cite{Booth+2017}.
However, this change in the disc gas composition would hardly affect the total mass of the water produced in the primordial atmosphere when the atmosphere is equilibrated with the magma ocean.
Secondly, pebbles drifting from outside the snowline would be more oxidised than solids originally existing inside the snowline~\cite{Lodders2003}.
Since the equilibrium partial pressure of the water vapour in the atmosphere is higher for more oxidised magma, the accretion of the oxidised rocky components would work in favour of the water production and the formation of planets with small water content.
Finally, a pebble accretion scenario may significantly change the resultant $M_\text{p}$-$a$ distribution of the synthesised planets~\cite{Brugger+2020}.
Since the water amount in the atmosphere largely depends on the planetary mass, the mass distribution of the planets in the habitable zone affects the water amount distribution.
Further studies including pebbles are needed to quantify these effects.

Our models place constraints on the radii of temperate 
aqua planets, which will be useful for ongoing and future exploration of habitable planets around M dwarfs.
Figure~\ref{fig:Rp_Fwater} shows the relationships between the radius and water content of HZ-NEMPs with enriched atmospheres orbiting $0.3M_\odot$ M dwarfs.
Most of the HZ-NEMPs of $> 1.3 R_\oplus$ are shown in bluish colours, indicating that those planets have thick hydrogen-rich atmospheres ($\gtrsim 10^{-3} M_\mathrm{p}$). 
Given those thick atmospheres bring about such a strong blanketing effect that an H$_2$O layer below the atmosphere would be, if any, in a super-critical state, 
those planets are unlikely to be habitable.
The other HZ-NEMPs of $> 1.3 R_\oplus$ (black symbols) retain large amounts of water ($\gtrsim 0.1 M_\mathrm{p}$) and 
are far from Earth-like (but worth atmospheric characterisation with JWST~\cite{Gardner+2006} and Ariel \cite{Tinetti+2018} for verifying our theoretical prediction).

The HZ-NEMPs of 0.7--1.3~$R_\oplus$ (shown in black) have lost their hydrogen atmospheres completely, ending up with rocky planets covered with oceans.
It turns out that those planets are diverse in water content and do include planets with Earth-like water content.
Several climate studies argue the amounts of seawater appropriate for temperate climates, considering the effects of seafloor weathering, high-pressure ice, water cycling and heterogeneous surface water distribution \cite{Abe+2011,Abbot+2012,Kaltenegger+2013,Alibert2014,Nakayama+2019,Moore+Cowan2020}. According to those studies, the appropriate seawater amount ranges from $\sim$0.1 to 100 times that of the Earth.
From an observational point of view, it would be important to exclude planets unlikely to be habitable in advance. Among those HZ-NEMPs, there are relatively low-mass rocky planets that have deep oceans with high-pressure ice; such planets are unlikely to have temperate climates~\cite{Kaltenegger+2013,Alibert2014,Nakayama+2019}. 
Meanwhile, one can identify rocky planets with ocean mass fractions $\gtrsim$~100 times the present-day Earth's one, which are also likely uninhabitable,
if planetary masses and radii are measured within 
$\lesssim$ 20\% and 5\% accuracies, respectively.
In Fig.~\ref{fig:Rp_Fwater}, about 25~\% of the HZ-NEMPs of 0.7--1.3~$R_\oplus$ are such ocean-dominated planets.

The remaining 75\% would be identified as ``water-poor rocky planets'' (i.e., rocky planets without hydrogen-rich atmospheres nor thick oceans), 
which are capable of having Earth-like temperate climates as long as they have little amounts of seawater~\citep[$\gtrsim$ 0.001\%;][]{Abe+2011,Kodama+2019,Moore+Cowan2020}.
Thus, excluding the completely dry planets, which account for about 95\% of those remaining HZ-NEMPs, the HZ-NEMPs with appropriate amounts of seawater for habitability are estimated to account for $\sim$5~\% of the ``water-poor rocky planets'' orbiting $0.3M_\odot$ M dwarfs.
This frequency becomes higher for larger stellar mass, and around $0.5M_\odot$ stars, for example, more than 10~\% of the water-poor rocky planets are expected to have the appropriate amounts of seawater.
Note that the appropriate water amount for temperate climate can be even wider, considering the possibility of the water sequestration in the mantle~\cite{Cowan+Abbot2014} and of further water loss due to the absence of silicate weathering~\cite{Abbot+2012}. On the other hand, for tidally locked planets, water can be trapped as ice on the night side, leading to atmospheric collapse, if the planet has only a small amount of the atmosphere~\cite{Joshi1997,Wordsworth+2011}. 
In this case, a temperate climate would be difficult to achieve.
Whether the planets with seawater contents focused on in this study can actually sustain a temperate climate needs to be addressed in the future.

Finally, the significance of the above occurrence rate of aqua planets with temperate climates may be realised by comparison with the number of HZ-NEMPs possibly detected by ongoing and future survey missions such as TESS and PLATO. TESS is estimated to detect $\sim$10 HZ-NEMPs around early M dwarfs of 0.3-0.5~M$_\odot$~\cite{Kunimoto+2022}, and PLATO will detect $\sim$100 HZ-NEMPs~\cite{Rauer+2014}. 
Multiplying the number of possible detections by the predicted occurrence rate, our model predicts that 5-10 aqua planets with temperate climates will be discovered in the 2020s, whereas the previous model predicting the occurrence rate of < 0.01\% \cite{Tian+Ida2015} suggests that the discovery is hopeless.

\section*{Method}

Here we describe the details about our planetary population synthesis model, which is based on the planetesimal-driven core accretion scenario.
The model includes the evolution of the central star, 
the evolution and radial structure of the protoplanetary disc,
the growth of solid cores,
the accumulation and loss of atmospheres, 
orbital migration,
and dynamical interactions between protoplanets,
including resonance trapping,
and orbital instabilities for multi-protoplanet systems and their outcomes.
In addition, we describe the calculation method for the radius of the isolated planet during its thermal evolution 
and the initial conditions and parameters for generating the planetary populations.
We discuss the effects of the updated components of the model from the previous study~\cite{Tian+Ida2015} and the consistency with the observed exoplanet populations in the Section 2.1 and 2.2 in the Supplementary Information, together with Supplementary Figs. 3 and 4.

\subsection*{Stellar Evolution} \label{sec:star}
From the table provided by ref.\cite{Baraffe+1998}, we take the stellar radius $R_*$, luminosity $L_*$ and effective temperature $T_*$
as a function of stellar mass $M_*$ and age for the solar metallicity with He abundance of $Y=0.275$. 
Those data are used for calculating the protoplanetary disc temperature, planetary equilibrium temperature, and atmospheric photoevaporation rate in our planetary population synthesis calculations.

\subsection*{Protoplanetary Disc} \label{sec:disc}
Our planetary population synthesis calculations need
the surface density distributions of gas and solids (or planetesimals) in the protoplanetary disc and their evolution, and also need the disc midplane temperature. 
Those are calculated in a similar way to ref. \cite{Emsenhuber+2021a}, which is summarised below.

\subsubsection*{Gas disc profile and evolution}
\begin{center}
\textit{(i) Initial profiles}
\end{center}

The initial profile of disc gas surface density is 
assumed as~\citep{Veras+Armitage2004,Andrews+2010}
\begin{align}
\begin{split}
    \Sigma_\text{g}^{(t=0)} &= \Sigma_\text{g0}\qty(\frac{r}{r_0})^{-q_\text{g}}
    \exp[ -\qty(\frac{r}{r_\text{disc}})^{2-q_\text{g}} ] \\
    & \qquad 
    \times \qty( 1 - \sqrt{\frac{r_\text{in}}{r}}),
    \label{eq:Sigmag0}
\end{split}    
\end{align}
where $\Sigma_\text{g0}$ is the gas surface density at orbital radius $r$ = $r_0$, 
$r_\text{disc}$ is the gas disc 
characteristic radius beyond which the surface density decays exponentially,
and $r_\text{in}$ is the inner edge radius.
This is the so-called self-similar solution with the turbulent viscosity $\nu_\text{acc} \propto r^{q_\text{g}}$~\citep{Lynden-Bell+Pringle1974,Hartmann+1998}, including the smooth cutoff at $r=r_\text{in}$.
We set $q_\text{g}=0.9$, which is inferred from observations of protoplanetary discs~\citep{Andrews+2010}, and $r_0=1$~au.
The surface density at $r$ = $r_0$, $\Sigma_\text{g0}$, is calculated from the total disc gas mass $M_\text{disc}$ as
\begin{equation}
    \Sigma_\text{g0} = \frac{2-q_\text{g}}{2\pi} M_\text{disc}
    r_0^{-q_\text{g}} r_\text{disc}^{q_\text{g}-2}.
\end{equation}
This is obtained by integrating Eq.~\eqref{eq:Sigmag0} multiplied by $2\pi r$ (i.e., $2\pi r \Sigma_\text{g}^{(t=0)}$) from $r=0$ to $r_\text{disc}$, ignoring the exponential and inner edge cutoff terms.

\begin{center}
\textit{(ii) Viscous diffusion} 
\end{center}

The evolution of the disc gas surface density $\Sigma_\text{g}$ is assumed to occur via radial viscous diffusion, photo-evaporation, and absorption by embedded planets and is expressed as~\citep[e.g.,][]{Lynden-Bell+Pringle1974}
\begin{equation}
    \pdv{\Sigma_\text{g}}{t} - \frac{3}{r}
    \pdv{r}( r^{1/2} \pdv{r}(\nu_\text{acc} \Sigma_\text{g} r^{1/2}))
    = -\dot{\Sigma}_\text{pe} - \dot{\Sigma}_\text{planet}
    \label{eq:dSigma_dt_basic},
\end{equation}
where $-\dot{\Sigma}_\text{pe}$ and $-\dot{\Sigma}_\text{planet}$ are the 
sink terms due to photo-evaporation and absorption by planets, respectively.
We adopt the $\alpha$-prescription for the turbulent viscosity:
$\nu_\text{acc}=\alpha_\text{acc} c_\text{s}H_\text{disc}$ \citep{Shakura+Sunyaev1973},
where $c_\text{s} = \sqrt{k_\text{B}T_\text{disc}/(\mu m_\text{H})}$ is the isothermal sound speed 
and $H_\text{disc}=c_\text{s}/\Omega_\text{K}$ is the disc scale height.
Here $\mu$ is the mean molecular weight, which is assumed to be 2.34 for gas with solar abundances, $T_\text{disc}$ is the midplane temperature calculated below, $k_\text{B}$ and $m_\text{H}$ are the Boltzmann constant ($=\SI{1.38e-16}{erg.K^{-1}}$) and the hydrogen atomic mass ($=\SI{1.66e-24}{g}$), respectively, and $\Omega_\text{K}$ is the Keplerian frequency. We treat $\alpha_\mathrm{acc}$ as an input parameter.

Equation~\eqref{eq:dSigma_dt_basic} is solved with a log-uniform grid with 
500 points ($N_\mathrm{grid,disc}$) extending from $r_\text{in}$ to $r_\text{max}=1000$~au.
The boundary conditions are $\Sigma_\text{g}(r_\text{in})=0$ and $\Sigma_\text{g}(r_\text{max})=0$.

\begin{center}
\textit{(iii) Photo-evaporation}
\end{center}

We consider photo-evaporation processes caused by UV irradiation from the central star (internal photo-evaporation) and from nearby massive stars (external photo-evaporation).
The total photo-evaporation rate $\dot{\Sigma}_\text{pe}$ is determined by
the sum of the two photo-evaporation rates. 
The external photo-evaporation rate is calculated by the prescription from ref. \cite{Matsuyama+2003}.
The model assumes that the FUV photons from nearby massive stars evaporate the disc gas uniformly only in the regions exterior to the effective gravitational radius.
The gravitational radius for the dissociated gas is defined as
\begin{equation}
    r_\text{g,I} = \frac{GM_*}{c_\text{s,I}^2},
\end{equation}
where $G$ is the gravitational constant ($=\SI{6.67e-8}{cm^3.g^{-1}.s^2} $);  $c_\text{s,I}$ is the sound speed with temperature of $1 \times 10^3$~K and mean molecular weight of 
1.35~\citep{Matsuyama+2003}, given that the hydrogen is fully-dissociated and the He/H ratio is equal to  solar.
Assuming that the disc gas is uniformly removed only from the regions exterior to the effective gravitational radius $\beta_\text{I}r_\text{g,I}$, the external photo-evaporation rate is given by 
\begin{equation}
    \dot{\Sigma}_\text{pe,ext} = 
    \begin{cases}
      0  &\qfor r < \beta_\text{I}r_\text{g,I} \\
\displaystyle{
      \frac{\dot{M}_\text{wind}}{\pi(r_\text{max}^2-\beta_\text{I}^2r_\text{g,I}^2)}}  & \quad \text{otherwise},
    \end{cases}
    \label{eq:dSigma_pe_ext}
\end{equation}
where we assume $r_\text{max}=1000$~au and regard $\dot{M}_\text{wind}$ as an input parameter providing the total mass loss rate.
For the effective gravitational radius, analytical estimates~\citep[e.g.,][]{Liffman2003} and numerical results~\cite[e.g.,][]{Begelman+1983a,Adams+2004,Font+2004} show that $\beta_\text{I}=$0.1--0.2 would be appropriate.
Here we set $\beta_\text{I}=0.14$ following ref.~\cite{Emsenhuber+2021a}.

Next, the internal photo-evaporation rate is calculated 
from the following equation \citep[e.g.,][]{Hollenbach+1994,Clarke+2001}
\begin{equation}
    \dot{\Sigma}_\text{pe,int}(r) = 
    \begin{cases}
      0  &\qfor r < \beta_\text{II}r_\text{g,II} \\
      2 c_\text{s,II}n_0(r) m_\text{H} & \quad \text{otherwise}.
    \end{cases}
\end{equation}
Here, $r_\text{g,II}$ and $c_\text{s,II}$ are the gravitational radius and sound speed for ionised gas of temperature $1 \times 10^{4}$~K and mean molecular weight 0.68~\citep[assuming the fully ionised hydrogen and the He/H ratio of solar;][]{Hollenbach+1994,Liffman2003,Matsuyama+2003}, respectively.
We also set $\beta_\text{II}=0.14$, as well as $\beta_\text{I}$.
The number density of the ionised hydrogen $n_0$ is 
given by
\begin{equation}
    n_0(r) = \num{3.1e5} \qty(\frac{r_\text{g,II}}{1~\si{au}})
    \qty(\frac{\Phi}{10^{40}~\si{s^{-1}}})^{1/2}
    \qty(\frac{r}{1~\si{au}})^{-5/2},
    \label{eq:n0}
\end{equation}
following the fitting formula derived in ref. \cite{Hollenbach+1994}, with $\Phi$ the ionising EUV photon luminosity.
Finally, the total photo-evaporation rate is given by
\begin{equation}
    \dot{\Sigma}_\text{pe} = \dot{\Sigma}_\text{pe,ext} + \dot{\Sigma}_\text{pe,int}.
\end{equation}

\begin{center}
\textit{(iv) Absorption by planets}
\end{center}

The sink term for absorption of disc gas by planets $\dot{\Sigma}_\text{planet}$ is simply given by
\begin{equation}
    \dot{\Sigma}_\text{planet} = \frac{\dot{M}_\text{env}}{4\pi a_\text{p} R_\text{H}},
    \label{eq:dSigma_planet}
\end{equation}
where $\dot{M}_\text{env}$ is the runaway gas accretion rate of the planet given by Eq.~\eqref{eq:gas_acc_rate}, and $a_\text{p}$ and $R_\text{H}$ are the semi-major axis and Hill radius of the planet, respectively.
Equation~\eqref{eq:dSigma_planet} is calculated for grids inside the Hill radius of each planet.

\subsubsection{Disc temperature}\label{sec:disk_temp}
The disc midplane temperature $T_\text{disc}$ is determined by the combination of viscous heating and direct and indirect stellar irradiation and can be expressed as~\citep[e.g.,][]{Emsenhuber+2021a}
\begin{equation}
    \sigma T_\text{disc}^4 = \sigma T_\text{vis}^4 + \sigma T_\text{irr}^4
    + \sigma T_\text{eq}^4 \exp(-\tau_r),
\end{equation}
where $T_\mathrm{vis}$ and $T_\mathrm{irr}$ are the temperatures that would be achieved if only viscous heating or indirect stellar irradiation, respectively, was available, and $T_\text{eq}$ is the equilibrium temperature due to the direct stellar irradiation through the disc midplane, and $\tau_r$ is the radial optical depth at the disc midplane.

The temperature $T_\text{vis}$ is given by \citep[e.g.,][]{Nakamoto+Nakagawa1994,Hueso+Guillot2005}
\begin{equation}
    \sigma T_\text{vis}^4 = \frac{1}{2}\qty(\frac{3}{8}\tau_\text{R} + \frac{1}{2\tau_\text{P}})\dot{E},
    \label{eq:Tdisc_NN}
\end{equation}
where $\tau_\text{R} (= \kappa_\text{R}\Sigma_\text{g})$ and $\tau_\text{P}$ are the Rosseland and Planck mean optical depths, respectively, and $\dot{E}$ is the viscous energy generation rate per unit surface area.
The Rosseland mean opacity $\kappa_\text{R}$ is given by the analytical fitting of the opacity of dust grains provided by ref. \cite{Bell+Lin1994}.
We ignore the gas opacity and simply set $\kappa_\text{R} = 0~\si{cm^2/g}$ after the evaporation of dust grains.
Also, we simply set $\tau_\text{P} = \max(2.4\tau_\text{R},0.5)$
\citep{Nakamoto+Nakagawa1994,Hueso+Guillot2005,Suzuki+2016}.
The minimum value of 0.5 is adopted so that the coefficient of $\dot{E}$ in Eq.~\eqref{eq:Tdisc_NN} approaches to unity in the optically thin limit.
The viscous heating rate $\dot{E}$ is 
\begin{equation}
    \dot{E} = \frac{9}{4}\nu_\text{acc} \Sigma_\text{g} \Omega_\text{K}^2.
\end{equation}

The temperature due to indirect irradiation $T_\text{irr}$ is given by \citep{Kusaka+1970,Adams+1988,Ruden+Pollack1991}
\begin{align}
\begin{split}
    T_\text{irr}^4 &= T_*^4 
    \left[ \frac{2}{3\pi}\qty(\frac{R_*}{r})^3 \right. \\
    & \qquad \left. + \frac{1}{2}\qty(\frac{R_*}{r})^2
    \frac{H_\text{disc}}{r}\qty(\dv{\ln H_\text{disc}}{\ln r}-1)\right].
    \label{eq:Tirr}
\end{split}    
\end{align}
For computational simplicity, we adopt a constant value of $\dv*{\ln H_\text{disc}}{\ln r}=9/7$, which is the approximate equilibrium solution for a flaring disc \citep{Chiang+Goldreich1997}.
Finally, $T_\text{eq}$ and $\tau_r$ are given by
\begin{equation}
    T_\text{eq}^4 = \frac{L_*}{16\pi \sigma r^2},
    \label{eq:Teq}
\end{equation}
and 
\begin{equation}
    \tau_r = \int_0^r \kappa_\text{R} \rho_\text{disc} \dd{r},
\end{equation}
where $\rho_\text{disc} = \Sigma_\text{gas}/\sqrt{2\pi}H_\text{disc}$ is the gas density at the disc midplane.

Since all of the 
temperatures above change with time, the H$_2$O snowline, defined as the radial location for 
$T_\text{disc}=170$~K, 
moves in the protoplanetary disc as follows.
In the early stage of the disc evolution, the disc midplane temperature $T_\text{disc}$ is almost 
equal to $T_\text{vis}$ and generally much higher than $T_\text{eq}$ because of the large viscous heating rate.
Thus, the snowline locates farther than the present position.
As the disc evolves, $T_\text{disc}$ becomes comparable to $T_\text{irr}$, which is much smaller than $T_\text{vis}$, and the snowline moves inward on a $\sim$Myr timescale.
Then, at the timing of the disc gas dissipation, $T_\text{disc}$ jumps up to $T_\text{eq}$ in a quite short time (typically $\sim 10^{4}$~yrs).
Since $T_\text{eq} > T_\text{irr}$ holds in many cases, the snowline moves outward in this stage.
Finally, the snowline slightly moves according to the evolution of the stellar luminosity (generally inward around 
M dwarfs).

\subsubsection{
Distribution and evolution of planetesimals}
We set the initial distribution of planetesimals (or solids) in terms of 
surface density as 
\begin{equation}
    \Sigma_\text{s}^{(t=0)} = \eta_\text{ice} \Sigma_\text{rock}^{(t=0)},
\end{equation}
where $\eta_\text{ice}$ is the enhancement factor associated with H$_2$O condensation (see below) and $\Sigma_\text{rock}^{(t=0)}$ is the initial surface density of rocks, which is given by
\begin{equation}
    \Sigma_\text{rock}^{(t=0)} = \Sigma_\text{s0}\qty(\frac{r}{r_0})^{-q_\text{s}}
     \exp\left[ -\qty(\frac{r}{r_\text{solid}})^2 \right]
    \qty( 1 - \sqrt{\frac{r_\text{in}}{r}});
    \label{eq:Sigma_solid}
\end{equation}
$\Sigma_\text{s0}$ is the 
initial surface density of rocks at $r=r_0$, 
and $r_\text{solid}$ is the 
characteristic radius of the existence region of planetesimals (or called the planetesimal disc).
We set $q_\text{s}=1.5$ \citep{Birnstiel+2012,Birnstiel+Andrews2014} and $r_\text{solid}=0.5r_\text{disc}$ \citep{Ansdell+2018}.
Similarly to $\Sigma_\text{g0}$, $\Sigma_\text{s0}$ is calculated by
\begin{equation}
    \Sigma_\text{s0} = \frac{2-q_\text{s}}{2\pi \eta_\text{ice}} M_\text{solid}
    r_0^{-q_\text{s}} r_\text{solid}^{q_\text{s}-2}.
\end{equation}
Here $M_\text{solid}$ is the total mass of planetesimals initially existing in the entire protoplanetary disc, which 
is given by
\begin{equation}
    M_\text{solid} = 10^{\text{[Fe/H]}}Z_{\odot}M_\text{disc},
\end{equation}
where [Fe/H] is the disc metallicity relative to the solar one, which is assumed equal to the central star's one, and $Z_\odot$ is the solar metallicity, which is set to 0.015~\citep{Lodders2003}.
The enhancement factor $\eta_\text{ice}$ is given by
\begin{equation}
    \eta_\text{ice} = 
    \begin{cases}
     1 & \qfor r < r_\text{ice}, \\
     2 & \qfor r > r_\text{ice},
    \end{cases}
    \label{eq:etaice}
\end{equation}
where $r_\text{ice}$ is the snowline position.
This is similar to the prescription in \cite{Ida+2013}, but we assume $\eta_\text{ice}=2$ beyond the snowline \citep{Lodders2003}, instead of 4.2 in their original prescription. We ignore snowlines for condensates other than H$_2$O.

The planetesimal distribution changes due to the accretion and scattering by protoplanets, and also due to the sublimation of ice as the snowline moves.
Assuming 
that $\Sigma_\text{rock}$ is only affected by the first two processes, we calculate its evolution by
\begin{equation}
    \dv{\Sigma_\text{rock}}{t} = -\frac{\dot{M}_\text{core}+\dot{M}_\text{scat}}{2\pi \eta_\text{ice} a_\text{p} \Delta a_\text{FZ}},
    \label{eq:dSigma_rock}
\end{equation}
where $\dot{M}_\text{core}$ and $\dot{M}_\text{scat}$ are the planetesimal accretion and scattering rates by the protoplanet, respectively, and $\Delta a_\text{FZ}$ is the full-width of the feeding zone of the planet, which is assumed to be 10~$R_\mathrm{H}$ \citep{Kokubo+Ida2002}.
The prescriptions for $\dot{M}_\text{core}$ and $\dot{M}_\text{scat}$ are described in section~\ref{sec:solid_core_growth}.
To calculate Eq.~\eqref{eq:dSigma_rock}, we set a log-uniform grid, which is different from that for disc gas, with $N_\text{grid,solid}=1000$ extending from $r_\text{in}$ to $5r_\text{solid}$.
To evaluate $\Sigma_\text{s}(t)$, we first integrate Eq.~\eqref{eq:dSigma_rock} to derive $\Sigma_\text{rock}(t)$ for all grids inside the feeding zone of each planet.
If a grid is in the feeding zone of multiple planets, the right hand side of Eq.~\eqref{eq:dSigma_rock} is calculated for all the involved planets and the results are summed up.
At the same time, $\eta_\text{ice}(t)$ for each grid is evaluated using $T_\text{disc}(t)$.
Then, the planetesimal surface density is derived by $\Sigma_\text{s}(t) = \eta_\text{ice}(t)\Sigma_\text{rock}(t)$

\subsection*{Solid Core Growth} \label{sec:solid_core_growth}
We assume that protoplanetary solid cores grow by the accretion of planetesimals.
The mass growth rate is given by
\begin{equation}
    \dv{M_\text{core}}{t} = \Omega_\text{K} \bar{\Sigma}_\text{s} R_\text{H}^2 p_\text{col}
\end{equation}
with $\bar{\Sigma}_\text{s}$ the mean surface density of planetesimals in the planet's feeding zone and $p_\text{col}$ the collision probability of planetesimals;
$p_\text{col}$ is given by \citep{Inaba+2001}
\begin{equation}
    p_\text{col} = \min \qty(p_\text{col,med}, \qty(p_\text{col,high}^{-2}+p_\text{col,low}^{-2})^{-1/2} ),
\end{equation}
with
\begin{align}
    p_\text{col,high} &= \frac{1}{2\pi} \qty(\frac{R_\text{cap}}{R_\text{H}})^2
    \qty(
    \mathcal{F}(i_\text{plt}/e_\text{plt}) + \frac{6R_\text{H}\mathcal{G}(i_\text{plt}/e_\text{plt})}{R_\text{cap}\tilde{e}_\text{plt}^2}
    ), \\
    p_\text{col,med} &=  \frac{1}{4\pi \tilde{i}_\text{plt}} \qty(\frac{R_\text{cap}}{R_\text{H}})^2
    \qty(
    17.3 + \frac{232R_\text{H}}{R_\text{cap}}
    ),\\
    p_\text{col,low} &= 11.3\sqrt{R_\text{cap}/R_\text{H}}.
\end{align}
Here $i_\text{plt}$ and $e_\text{plt}$ are respectively the inclination and eccentricity of planetesimals, 
and $\tilde{i}_\text{plt}=i_\text{plt}/h$ and $\tilde{e}_\text{plt} = e_\text{plt}/h$ are the reduced inclination and eccentricity, respectively, with $h = R_\text{H}/a_\text{p}$.
We assume $i_\text{plt} = e_\text{plt}/2$, in which case $\mathcal{F}(i_\text{plt}/e_\text{plt}) = 17.3$ and $\mathcal{G}(i_\text{plt}/e_\text{plt}) = 38.2$
~\citep[see][for the explicit form of $\mathcal F(x)$ and $\mathcal G(x)$]
{Greenzweig+Lissauer1990,Greenzweig+Lissauer1992,Inaba+2001}.
The value of $\tilde{e}_\text{plt}$ around a protoplanet is assumed equal to the equilibrium eccentricity determined by the balance of viscous stirring by the planet and the damping due to the disc gas drag.
Following the analytic estimates by ref.\cite{Kokubo+Ida2002}, it is expressed as
\begin{align}
\begin{split}
     \tilde{e}_\text{plt}
    = 4.1 
    &\qty(\frac{\rho_\text{disc}}{\SI{1e-9}{g/cc}})^{-1/5}
    \qty(\frac{\rho_\text{plt}}{\SI{3}{g/cm^3}})^{2/15} \\
    & \times\qty(\frac{a_\text{p}}{1~\si{au}})^{-1/5}
    \qty(\frac{m_\text{plt}}{10^{20}\si{g}})^{1/15},
\end{split}    
    \label{eq:tilde_e_plt}
\end{align}
where $\rho_\text{plt}=3.0~\si{g.cm^{-3}}$ and $m_\text{plt}=10^{20}~\si{g}$ are the material density and mass of planetesimals.
Finally, 
we calculate the effective capture radius enhanced by the atmospheric gas drag, $R_\text{cap}$,
following ref.~\cite{Inaba+Ikoma2003};
\begin{equation}
    R_\text{plt} = \frac{3}{2}
    \frac{v_\text{ran}^2 + 2GM_\text{p}/R_\text{cap}}
    {v_\text{ran}^2 + 2GM_\text{p}/R_\text{H}}
    \frac{\rho(R_\text{cap})}{\rho_\text{plt}} R_\text{cap},
    \label{eq:Rcap}
\end{equation}
with $R_\text{plt}$ the planetesimal's radius derived from $m_\text{plt}$ and $\rho_\text{plt}$, $v_\text{ran} = e_\text{plt}a_\text{p}\Omega_\text{K}$ the random velocity of the planetesimals and $\rho(R_\text{cap})$ the envelope gas density at the radius of $R_\text{cap}$.

After the disc gas dissipation, some of the planetesimals that encountered a planet are ejected from the system, instead of colliding with the planet.
This ejection rate can be estimated by comparing the collisional cross section and the scattering cross section, and is given by~\citep{Ida+Lin2004}
\begin{equation}
    \dot{M}_\text{scat} = \qty(\frac{v_\text{surf}}{v_\text{esc}})^4 \dot{M}_\text{core},
\end{equation}
where $v_\text{esc}=\sqrt{2GM_*/a_\text{p}}$ is the escape velocity from the host star and $v_\text{surf}=\sqrt{GM_\text{p}/R_\text{cap}}$ is the surface velocity of the planet.
This is used for the calculation of the evolution of $\Sigma_\text{rock}$ (Eq.~\eqref{eq:dSigma_rock}).

\subsection*{Atmospheric Accumulation and Loss} \label{sec:atmosphere}
\subsubsection*{Purely hydrostatic equilibrium phase}
During vigorous planetesimal accretion, the atmospheric mass is rather small because of significant energy deposition. Then, the atmosphere is in the hydrostatic equilibrium and thermally steady state \citep[e.g.,][]{Ikoma+2000}. We solve the standard set of equations for stellar structure, namely,
\begin{align}
    \pdv{P}{M_R} &= -\frac{GM_R}{4\pi R^4}, \label{eq:dPdm} \\
    \pdv{T}{M_R} &= -\frac{GM_R}{4\pi R^4}\frac{T}{P} \nabla , \label{eq:dTdm} \\
    \pdv{R}{M_R} &= \frac{1}{4\pi R^2 \rho} \label{eq:drdm}, 
\end{align}
where $P$, $T$, and $\rho$ are the pressure, temperature, and density of the atmosphere (or envelope) gas, respectively, $R$ is the radial distance to the planet's centre, and $M_R$ is the total mass inside the sphere of radius $R$.
In addition to the above equations, we use the ideal equation of state (or the $P$-$T$ relationship) for chemical equilibrium mixtures composed of H- and O-bearing molecules and He, taking H$_2$O sublimation/condensation into account \citep[see][for the details]{Kimura+Ikoma2020}. Also, $\nabla$ is the temperature gradient, $\dv*{\log T}{\log P}$, for radiative diffusion or convection (dry or moist adiabat).

In the early stages of planetary accretion, the temperature is high enough at the bottom of the atmosphere to keep 
a global magma ocean~\citep[][]{Ikoma+Genda2006,Kimura+Ikoma2020}. The atmosphere-magma interaction produces volatiles, modifying the atmospheric composition significantly.
Therefore, we assume that the water-producing reaction effectively occurs and that the produced water vapour is uniformly mixed in the atmosphere.
Hereafter, we call such an atmosphere at this phase the \textit{vapour-mixed atmosphere}, 
and denote its mass by $M_\text{mix}$.

We calculate the atmospheric mass $M_\mathrm{mix}$ in the same way as ref.~\cite{Kimura+Ikoma2020}, treating
the water mass fraction in the vapour-mixed atmosphere $X_{\text{H}_2\text{O}}$ as an input parameter. 
We set $T = T_\text{disc}$ and $P = P_\text{disc}$ at $R = \min(R_\text{B}, R_\text{H})$, where $R_\text{B}$ and $R_\text{H}$ are the Bondi and Hill radii, respectively.
The luminosity $L$, which is assumed constant in the entire atmosphere, is equal to that from the solid core; namely,
\begin{equation}
    L_\text{core} = L_\text{acc} + L_\text{cool} + L_\text{radio},
    \label{eq:Lcore}
\end{equation}
where $L_\text{acc}$, $L_\text{cool}$ and $L_\text{radio}$ are the luminosity due to the planetesimal accretion, the solid core cooling and radioactive decay, respectively. The accretion luminosity
\begin{equation}
    L_\text{acc} = \frac{G M_\text{core} \dot{M}_\text{core}}{R_\text{core}},
    \label{eq:Lacc}
\end{equation}
where 
$\dot{M}_\text{core}$ is the planetesimal accretion rate, and $R_\text{core}$ is the solid planet radius.
By assuming that the core surface temperature $T_\text{surf}$ and the disc gas temperature $T_\text{disc}$ are related in the form of $\dv*{T_\text{surf}}{\rho_\text{disc}}=T_\text{disc}/4$, as indicated by the analytical solution of the fully-radiative atmosphere, 
$L_\text{cool}$ is given by~\citep{Ikoma+Hori2012}
\begin{equation}
    L_\text{cool} = \frac{M_\text{core} C_\text{rock} T_\text{disc}}{4\tau_\text{disc}},
    \label{eq:Lcool}
\end{equation}
where $C_\text{rock}$ is the specific heat of rock $(=\SI{1.2e7}{erg \, g^{-1}K^{-1}})$, and $\tau_\text{disc}$ is the disc dissipation timescale.
We set $\tau_\text{disc}$ = $1 \times 10^4$~yr in our simulations, since disc dissipation occurs on a timescale of $\sim 10^4$--$10^5$~yr due to photoevaporation in the final stage of disc evolution.
Increasing this value by an order of magnitude has little effect on our results.
While the above equation was derived assuming a thin radiative atmosphere, the vapour-rich primordial atmosphere considered in this study is thick even in the final stage of the disc evolution, and the lower layer is often convective. 
Even in that case, the changing rate of temperature at the bottom of the atmosphere is about the same as that at the radiative-convective boundary.
Therefore, Eq.~\eqref{eq:Lcool} is still considered to be a good approximation for describing the cooling luminosity.
The radiogenic luminosity $L_\text{radio}$ is simply set to $\num{2e20}(M_\text{core}/M_\oplus)~\si{erg/s}$~\citep{Guillot+1995}.

\subsubsection*{Quasi-static contraction phase}
Once planetesimal accretion stops, 
the vapour-mixed atmosphere contracts gravitationally and further accumulation of disc gas occurs. 
Thus, we have to integrate the equation of entropy change, in addition to Eqs.~(\ref{eq:dPdm})-(\ref{eq:drdm});
\begin{equation}
        \pdv{L_R}{M_R} = \dot{\varepsilon} -T \dv{S}{t}, 
        \label{eq:dLdm}
\end{equation}
where $L_R$ is the total energy flux passing through a spherical surface of radius $R$ (or luminosity), $\dot{\varepsilon}$ is the energy generation rate per unit mass, and $S$ is the specific entropy.
We assume that the accumulating disc gas composed predominantly of H and He never mixes with the lower vapour-mixed layer and, thus, the mass of the vapour-mixed atmosphere ($M_\text{mix}$) is conserved in this phase.
Hereafter, the accumulated mass of disc gas is denoted by $M_\text{HHe}$.
Therefore we calculate the quasi-static contraction of the two-layer atmosphere in this phase.

The numerical integration of Eq.~(\ref{eq:dLdm}) is done based on the total energy conservation approximation adopted in several previous studies on the formation of gas giants~\citep{Papaloizou+Nelson2005,Mordasini+2012b,Fortier+2013,Piso+Youdin2014,Venturini+2016}.
The total atmospheric energy conservation between the time $t-\Delta t$ and $t$ being considered, the following relation holds (see ref.~\cite{Piso+Youdin2014} for the derivation):
\begin{align}
\begin{split}
    & \frac{E_\text{env}(t)-E_\text{env}(t-\Delta t)}{\Delta t} \\
    &\qquad = L_\text{core}
    + e_\text{gas} \frac{M_\text{env}(t) - M_\text{env}(t-\Delta t)}{\Delta t} - L
\end{split}
    \label{eq:energy_conserv}
\end{align}
or
\begin{equation}
    \Delta t = 
        \frac{E_\text{env}(t)-E_\text{env}(t-\Delta t)-e_\text{gas} \left[M_\text{env}(t) - M_\text{env}(t-\Delta t)\right]}
        {L_\mathrm{core}-L}
    \label{eq:energy_conserv2}
\end{equation}
where $e_\text{gas}$ is 
the total energy per unit mass (i.e., the sum of the specific internal energy and the gravitational energy)
of the disc gas at the outer boundary, $M_\text{env} = M_\text{mix}+M_\text{HHe}$ 
is the total envelope mass, and $E_\text{env}$ is the envelope's total energy defined by
\begin{equation}
    E_\text{env} = \int_{M_\text{core}}^{M_\text{p}} \qty( u - \frac{GM_R}{R}) \dd{M_R};
\end{equation}
$u$ is the specific internal energy.
In Eq.~\eqref{eq:energy_conserv}, we have neglected the work done by the atmospheric surface (i.e. the boundary between the atmosphere and disc gas) for simplicity.
This term generally accounts for only a few \% relative to the other terms \citep[e.g.,][]{Lee+2014} and hardly affects the results of this study.

Once planetesimal accretion is over, the luminosity decreases with the mass growth of the envelope until reaching the critical luminosity below which no hydrostatic solution is found \citep[][]{Ikoma+2000,Hubickyj+2005}. Thus, for $L (t)$ = $\max [0.95 L (t-\Delta t), L_\mathrm{core}]$, we integrate Eqs.~(\ref{eq:dPdm})-(\ref{eq:drdm}) to calculate
$E_\text{env}(t)$ and  $M_\text{env}(t)$, assuming $L_R$ = $L (t)$, and thereby calculate $\Delta t$ from Eq.~(\ref{eq:energy_conserv2}).
The assumption of $L_R = L$ is valid until the critical luminosity while invalid after that \citep[][]{Ikoma+2000}.
The H$_2$O fraction $X_{\text{H}_2\text{O}}$ is set to zero for $M_R >M_\text{core}+ M_\text{mix}$, and is set to the input value otherwise.
Once the critical luminosity is reached, we shift to the runaway gas accretion regime (see the following section).

\subsubsection*{Runaway gas accretion phase} \label{sec:runaway_gas_acc}
The runaway gas accretion phase is divided into three sub-phases \citep[e.g.,][]{Tanigawa+Ikoma2007}. In the earlier phase the gas accretion is regulated by the Kelvin-Helmholtz contraction of the atmosphere.
Numerical simulations show that the contraction timescale is strongly dependent on planet mass \citep[][]{Tajima+1997,Ikoma+2000}. 
Here we assume that the growth timescale, $\tau_\mathrm{KH}$, increases with the cube of planetary mass, following the previous planetary population synthesis models \citep[e.g.,][]{Ida+2013}.
 \begin{equation}
     \tau_\text{KH} = 1 \times 10^9 \qty(\frac{M_\text{p}}{M_\oplus})^{-3}~\si{yr}.
     \label{eq:tKH}
 \end{equation}
While \citet{Tajima+1997} showed a stronger dependence, our results are insensitive to the choice of the power index, provided the index is smaller than $-3$, partly because the duration of this phase is short.
When the planet grows large enough to open a gap in the surrounding disc, the growth is limited by the gas supply from the flow in the gap.
The supply rate is given by~\citep{Tanigawa+Ikoma2007,Tanigawa+Tanaka2016,Tanaka+2020}
 \begin{equation}
     \dv{M_\text{p}}{t} = D \Sigma_\text{g,gap},
     \label{eq:dMp_dt_gap}
 \end{equation}
where the coefficient $D$ is empirically given by
\begin{equation}
    D = 0.29\qty(\frac{H_\text{disc}}{a_\text{p}})^{-2}
        \qty(\frac{M_\text{p}}{M_*})^{4/3} a^2 \Omega_\text{K},
\end{equation}
and $\Sigma_\text{g,gap}$ is the gas surface density at the bottom of the gap, which is also empirically given by \citep{Kanagawa+2015}
\begin{equation}
    \Sigma_\text{g,gap} = \frac{\Sigma_\text{g}}{1+0.04K},
\end{equation}
with $K$ being 
\begin{equation}
    K = \qty(\frac{M_\text{p}}{M_*})^2 \qty(\frac{h}{a})^{-5}
        \alpha_\text{vis}^{-1}.
        \label{eq:Kanagawa}
\end{equation}
Here, $\alpha_\text{vis}$ is a parameter for disc turbulent viscosity, which is not necessarily the same as $\alpha_\text{acc}$.
When the wind driven accretion is dominant in the global angular momentum transfer, $\alpha_\text{acc}$ is larger than $\alpha_\text{vis}$ by about one order of magnitude~\citep{Simon+2013,Armitage+2013,Hasegawa+2017}.
Thus, we set $\alpha_\text{vis}=0.1\alpha_\text{acc}$ in this study.

Equation~\eqref{eq:dMp_dt_gap} is valid during there is sufficient disc gas outside the gap.
As the disc depletes, the gas accretion rate is limited by the global disc accretion rate $\dot{M}_\text{disc}$ given by 
\begin{equation}
    \dot{M}_\text{disc} = 6\pi r^{1/2}
    \pdv{r}(\Sigma_\text{g} \nu_\text{acc} r^{1/2}).
\end{equation}
Considering these processes, the runaway gas accretion rate is calculated by
\begin{equation}
    \dv{M_\text{p}}{t} = \min\qty(\frac{M_\text{p}}{\tau_\text{KH}}, D\Sigma_\text{g,gap}, \dot{M}_\text{disc}).
    \label{eq:gas_acc_rate}
\end{equation}
Note that the solid accretion never occurs in this phase.

\subsection*{Atmospheric Thermal Evolution and Loss} \label{sec:radius}

\subsubsection*{Atmospheric Thermal Evolution}
The planetary radius $R_\text{p}$ is expressed by
\begin{equation}
    R_\text{p} = R_\text{core} + \Delta R_\text{env},
\end{equation}
where $R_\text{core}$ is the solid core radius and $\Delta R_\text{env}$ is the thickness of the envelope. 

The solid core is assumed to consist of iron, silicate, and, if present, ice.
Following ref.\cite{Zeng+2019}, we calculate the core radius as
\begin{equation}
    R_\text{core}
    = R_\text{rock}(1+0.55 f_\text{ice} - 0.14f_\text{ice}^2),
\label{eq:Rcore}
\end{equation}
where $f_\text{ice}$ is the ice mass fraction and $R_\text{rock}$ is the pure rocky (iron+silicate) core radius, which is calculated from ref.~\cite{Fortney+2007} as
\begin{align}
\begin{split}
    R_\text{rock} 
    &= (0.0592f_\text{rock}+0.0975) (\log M_\text{core})^2 \\
    & \qquad + (0.2337f_\text{rock}+0.4938) \log M_\text{core} \\
    & \qquad + (0.3102f_\text{rock}+0.7932),
\end{split}    
    \label{eq:Rrock}
\end{align}
with $f_\text{rock}$ being the Si/(Si+Fe) mass ratio set to the Earth-like value of 0.66 in this study.

To evaluate the planetary radius including the envelope, we calculate the quasi-static thermal evolution of the planet after the disc gas dissipation.
For numerical convenience, we treat the upper part of the envelope (simply called the atmosphere below) separately from its deeper part (called the envelope), following previous evolution models of ice giant planets \citep[e.g.,][]{Kurosaki+Ikoma2017}.
The envelope structure is calculated by directly integrating Eqs.~\eqref{eq:dPdm}--\eqref{eq:drdm} and Eq.~\eqref{eq:dLdm} for a given water mass fraction $X_{\text{H}_2\text{O}}$.
For the equation of state, we use SCvH~\citep{Saumon+1995} for H-He and SESAME~\citep{Lyon+Johnson1992} for H$_2$O and mix H-He and H$_2$O according to the additive-volume law and the ideal mixing for entropy.

The inner boundary condition for the envelope structure is
\begin{equation}
    L_R = L_\text{core}, \quad R = R_\text{core} \qquad \text{at}\, M_R = M_\text{core}.
\end{equation}
Here $R_\text{core}$ is calculated with Eq.~\eqref{eq:Rcore}.
The core luminosity $L_\text{core}$ is the same as Eq.~\eqref{eq:Lcore} with $L_\text{acc}=0$.
We also calculate $L_\text{cool}$ self-consistently during the quasi-static evolution with
\begin{align}
\begin{split}
    L_\text{cool} &= -M_\text{core} C_\text{rock}\dv{T_\text{surf}}{t} \\
    &= -M_\text{core} C_\text{rock}\frac{T_\text{surf}(t) - T_\text{surf}(t-\Delta t)}{\Delta t}
\end{split}    
\end{align}
where $T_\text{surf}$ is the temperature at the solid core surface (i.e. the bottom of the envelope).
Since $T_\text{surf}(t)$ is determined by the envelope structure, we should find a self-consistent value of $L_\text{cool}$ iteratively.

The outer boundary condition for the envelope structure is given in terms of pressure and temperature, respectively, by
\begin{equation}
    P = P_\text{out}, \quad T = T_\text{out} \qquad \text{at}\, M_R = M_\text{p}.
\end{equation}
To evaluate $P_\text{out}$ and $T_\text{out}$, we calculate the radiative-convective structure of the atmosphere in the same way as~\cite{Kurosaki+Ikoma2017}.
The atmosphere is assumed to be plane parallel and its mass and thickness are negligible compared to the total planetary mass and radius, respectively.
Thus, the gravitational acceleration $g=GM_\text{p}/R_\text{out}^2$ is constant through the atmosphere, with $R_\text{out}$ denoting the radius at the boundary between the atmosphere and the envelope (note that it is not equal to the planetary radius $R_\text{p}$; see below).
This boundary is assumed to locate at the radius where the optical depth for the stellar visible radiation, $\tau_\text{vis}$, is 10,
so that the temperature gradient in the envelope is hardly affected by the irradiation from the central star.

The temperature in the radiative region is calculated by the analytical formula given by ref.~\cite{Matsui+Abe1986b};
\begin{equation}
\begin{split}
    \sigma T^4 &= F_\text{p} \frac{D\tau_\text{IR}+1}{2}\\
    & \quad+\frac{\sigma T_\text{eq}^4}{2}
    \qty[
    1 + \frac{D}{\gamma} + \qty( \frac{\gamma}{D} - \frac{D}{\gamma})
    \exp(-\tau_\text{vis})
    ],
\end{split}    
\end{equation}
where $\tau_\text{IR}$ is the optical depth for infrared radiation, $F_\text{p}=L_\text{p}/4\pi R_\text{out}^2$ is the net energy flux from the planet, with $L_\text{p}$ denoting the luminosity at the top of the envelope, $T_\text{eq}$ is the equilibrium temperature calculated with Eq.~\eqref{eq:Teq}, $D=3/2$ is the diffusivity factor arising from the angular dependence of the radiation flux, and $\gamma = \kappa_\text{vis}/\kappa_\text{IR}$ with $\kappa_\text{vis}$ and $\kappa_\text{IR}$ being the opacities for visible and infrared radiation, respectively.
We set $\gamma=0.1$ following \cite{Kurosaki+Ikoma2017}.

The radiative-convective boundary (i.e. tropopause) is determined so that the temperature and the radiation flux connects continuously, using the same numerical procedure as \cite{Kurosaki+Ikoma2017}.
As a result, $P_\text{out}$ and $T_\text{out}$ 
are determined when $L_\text{p}$ and $R_\text{out}$ are given from the envelope structure calculation.
Thus, we find a self-consistent set of $(P_\text{out}, T_\text{out}, L_\text{cool}, L_\text{p}, R_\text{out})$ at the time $t$ for a given $M_\text{p}$ with iterations.
Then we define the planetary radius $R_\text{p}$ as the pressure level of 10~mbar, and the envelope thickness is derived simply by $\Delta R_\text{env} = R_\text{p}-R_\text{core}$.

If the planet has experienced the runaway accretion phase, however, the thick envelope with nebular composition exists instead of the water-enriched envelope.
The radius in this case is calculated with the fitting formula given by \cite{Lopez+Fortney2014} as;
\begin{align}
\begin{split}
    \Delta R_\text{env} &= 2.06R_\oplus
    \qty(\frac{M_\text{p}}{M_\oplus})^{-0.21}
    \qty(\frac{f_\text{env}}{0.05})^{0.59} \\
    &\qquad \times \qty(\frac{F_\text{p}}{F_\oplus})^{0.044}
    \qty(\frac{t}{\si{yr}})^{-0.11} + \Delta R_\text{corr},
\end{split}
\end{align}
where $f_\text{env}=M_\text{env}/M_\text{p}$ and $F_\text{p}$ is the stellar insolation.
$\Delta R_\text{corr}$ is the correction factor defined as
\begin{equation}
    \Delta R_\text{corr} = \frac{9k_\text{B}T_\text{eq}}{g \mu_\text{neb}m_\text{H}},
\end{equation}
with $g$ the gravitational acceleration and $\mu_\text{neb}=2.34$ the mean molecular weight of the nebular gas.

\subsubsection*{Atmospheric Photoevaporation}

After the disc dispersal, the atmospheric escape (or photoevaporation) occurs because of stellar XUV irradiation.
We assume that the atmospheric photoevaporation starts once the radial optical depth at the planet's location measured from the central star, $\tau_r$, becomes less than unity.
We use the fitting formula by \cite{Kubyshkina+2018b,Kubyshkina+2018a} for the escape rate, which yields higher values of the escape rate than those of the energy-limited escape rate when the escape parameter is small ($\lesssim 10$).
Although the atmospheres in our calculations are enriched with water vapour, we simply adopt the formula derived for hydrogen-dominated atmospheres.
Since the enrichment with water leads to raising the value of the escape parameter and, thus, reducing the escape rate, our calculations are expected to overestimate the escape rate.
The fitting formula is expressed in the form of 
\begin{equation}
    \dot{M}_\text{esc} = e^\beta 
    \qty(\frac{F_\text{XUV}}{\si{erg.cm^{-2}.s^{-1}}})^{\alpha_1}
    \qty(\frac{a}{\si{au}})^{\alpha_2}
    \qty(\frac{R_\text{p}}{R_\oplus})^{\alpha_3}
    \Lambda^k,
\end{equation}
where $\beta, \alpha_1, \alpha_2, \alpha_3, k$ are the fitting coefficients~\citep[see][for their values]{Kubyshkina+2018a}, $F_\text{XUV}$ is the incoming XUV flux, and $\Lambda$ is the escape parameter.
The stellar XUV luminosity, $L_\text{XUV}$, is taken from the table given by \cite{Johnstone+2021}.
Since the table just gives the best-fit relation between the stellar $L_\text{XUV}$ and time, we set $L_\text{XUV} = 10^{\delta \log L_\text{XUV}} L_\text{XUV,bf}$ to account for the variation between observed stars, according with \cite{Johnstone+2021}.
Here, $L_\text{XUV,bf}$ is the best-fit value of XUV luminosity given in the table, and $\delta \log L_\text{XUV}$ is the deviation factor which follows the normal distribution with $\mu=0$ and $\sigma = 0.359$.

\subsection*{Orbital Migration} \label{sec:migration}

The planetary orbital migration rate can be calculated by
\begin{equation}
    \dv{a_\text{p}}{t} = \frac{2\Gamma}{M_\text{p} a_\text{p} \Omega_\text{K}},
    \label{eq:da_dt}
\end{equation}
where $\Gamma$ is the total torque acting on the planet.
In the type-I regime, the torque is given as the sum of the Lindblad ($\Gamma_\mathrm{L}$), corotation ($\Gamma_\mathrm{C})$, and thermal ($\Gamma_\mathrm{T}$) components:
\begin{equation}
    \Gamma = \Gamma_\text{L} + \Gamma_\text{C} + \Gamma_\text{T}.
    \label{eq:Gamma_typeI}
\end{equation}

We adopt the formula for the Lindblad torque derived by \cite{Jimenez+Masset2017} through 3D hydrodynamic simulations as
\begin{equation}
    \Gamma_\text{L} = (-2.34+1.50\beta_{T}-0.10\beta_{\Sigma})f(\chi_\text{P})\Gamma_0,
\end{equation}
where $\beta_{T}=\dv*{\log T_\text{disc}}{\log r}$, $\beta_{\Sigma}=\dv*{\log \Sigma_\text{g}}{\log r}$, $\chi_\text{P}$ is the thermal diffusion coefficient at the disc midplane, and $\Gamma_0$ is the characteristic torque expressed as
\begin{equation}
    \Gamma_0 = \qty(\frac{M_\text{p}}{M_*})^2 \qty(\frac{H_\text{disc}}{a_\text{p}})^{-2}\Sigma_\text{g} a_\text{p}^4 \Omega_\text{K}^2.
\end{equation}
The function $f(\chi_\text{P})$ transitions from $1/\gamma$ (`adiabatic regime') to unity (`locally isothermal regime') as $\chi_\text{P}$ increases (i.e. the thermal diffusion of the disc gas becomes effective).
Here $\gamma$ is the specific heat ratio of the disc gas.
See \cite{Jimenez+Masset2017} for the explicit form of $\chi_\text{P}$ and $f(\chi_\text{P})$.

We also use the formula for the corotation torque derived by \citet{Jimenez+Masset2017}, which is expressed as the sum of four components:
\begin{equation}
    \Gamma_\text{C} = \Gamma_\text{C,vor}+\Gamma_\text{C,ent}+\Gamma_\text{C,temp}+\Gamma_\text{C,vv},
\end{equation}
where $\Gamma_\text{C,vor}$, $\Gamma_\text{C,ent}$, and $\Gamma_\text{C,temp}$ are the torques arising from the radial gradient of vortensity, entropy, and temperature.
The last term $\Gamma_\text{C,vv}$ is associated with the viscously created vortensity arising during the horse-shoe U-turns~\citep[see][for more detailed explanations]{Masset+Casoli2009a,Masset+Casoli2009b,Masset+Casoli2010}.
Each of the first three torques is calculated by the combination of two terms; for example,
\begin{equation}
    \Gamma_\text{C,vor} = (1-\epsilon_b) \Gamma_\text{vor,lin} + \epsilon_b \Gamma_\text{vor,hs},
\end{equation}
where $\Gamma_\text{vor,lin}$ and $\Gamma_\text{vor,hs}$ are the vortensity-related components of the linear corotation torque and the horseshoe torque, respectively.
The horseshoe torque $\Gamma_\text{vor,hs}$ is the product of the unsaturated horseshoe drag $\Gamma_\text{vor,uhs}$ and the saturation function $\mathcal{F}_\text{v}$ (i.e., $\Gamma_\text{vor,uhs} \mathcal{F}_\text{v}$); the latter expresses the saturation of horseshoe drag and depends on horseshoe width.
The blending coefficient $\epsilon_b$ is derived in \cite{Masset+Casoli2010} by fitting their numerical results.
Note that the saturation function and the blending coefficient are different among $\Gamma_\text{C,vor}$, $\Gamma_\text{C,ent}$ and $\Gamma_\text{C,temp}$.
The final term $\Gamma_\text{C,vv}$ 
has only the component of the horseshoe torque (i.e., no linear corotation torque)
because this torque is only arising from the horse-shoe region.
Again, see \cite{Jimenez+Masset2017} for the explicit forms of these torques.

The thermal torque $\Gamma_\text{T}$ arises when the horseshoe gas cools during the U-turn by thermal diffusion (cold torque) and when the horseshoe gas is heated by the luminous planet (heating torque).
We calculate the thermal torque using the analytical formulation by \cite{Masset2017};
\begin{equation}
    \Gamma_\text{T} = 1.61\frac{\gamma-1}{\gamma}\eta \qty(\frac{H_\text{disc}}{\lambda_c})
    \qty(\frac{L_\text{p}}{L_\text{c}}-1)\Gamma_0,
\end{equation}
where $\eta = -\beta_{\Sigma}/3 - \beta_{T}/6 + 1/2$, $\lambda_c = \sqrt{\chi_\text{P}/(q\Omega_\text{K}\gamma)}$ with $q = M_\text{p}/M_*$, $L_\text{p}$ is the planetary luminosity,
and $L_\text{c} = 4\pi GM_\text{p}\chi_\text{P}\rho_\text{disc}/\gamma$ is the critical planetary luminosity above which the positive heating torque exceeds the negative cold torque.
Note that, as discussed in \cite{Masset2017}, the thermal torque acts effectively only when 
$M_\text{p} < \chi_\text{P}c_\text{s}/G$ is satisfied.
This condition means that the thermal diffusion timescale across the Bondi radius is shorter than the acoustic timescale.
If $M_\text{p}>\chi_\text{P}c_\text{s}/G$, we set $\Gamma_\text{thermal} = 0$.

When the planet grows massive enough to open a gap in the disc, its orbital migration transitions to the type-II regime.
Recently, \cite{Kanagawa+2018} found that the migration rate in this regime can be expressed in the similar way as that in type I regime by replacing gas surface density $\Sigma_\text{g}$ with $\Sigma_\text{gap}$ and removing the corotation torque.
To connect the both regimes smoothly, we express the total torque in the type II regime as
\begin{equation}
    \Gamma = \frac{\Gamma_\text{L} + (\Gamma_\text{C} + \Gamma_\text{T})\exp(-K/K_t)}
    {1+0.04K},
    \label{eq:Gamma_typeII}
\end{equation}
where $K_t=20$ is the typical value of $K$ where the disc gas gap becomes deep enough \citep{Kanagawa+2018}.
This is similar to the formula given in \cite{Kanagawa+2018}, 
but the formulae of $\Gamma_\text{L}$ and $\Gamma_\text{C}$ are different, and $\Gamma_\text{T}$ is also added.
Since Eq.~\eqref{eq:Gamma_typeII} becomes equal to Eq.~\eqref{eq:Gamma_typeI} when $K$ is small (i.e., the planet mass is small), we always use Eq.~\eqref{eq:Gamma_typeII} in the calculations of this study.

\subsection*{Resonance trapping} \label{sec:resonance}

Planets or planetary embryos with converging orbits can be captured in mean-motion resonances with each other; this phenomenon is called resonance trapping \citep{Murray+Dermott1999}.
In this study, the resonance trapping process is included in a similar way to refs. \cite{Ida+Lin2010} and \cite{Ida+2013}.
The treatment differs depending on whether the pair includes a giant planet or not---the definition of a `giant planet' is given in Section~\ref{sec:dyanamic_interact},
while an `embryo' refers to a celestial body other than the giant planet.

\subsubsection*{Case with two embryos}

When the orbits of two adjacent embryos (denoted by $i$ and $j$) are converging and get close enough to each other, their separation ($b = |a_i-a_j|$) expands impulsively at every conjunction.
The change in separation via the orbital repulsion is estimated by a linear analysis as~\citep{Goldreich+Tremaine1982,Hasegawa+Nakazawa1990}
\begin{equation}
    \Delta b = 30\qty(\frac{b}{r_\text{H}})^{-5} r_\text{H},
\end{equation}
where $r_\text{H} = ((M_i+M_j)/3M_*)^{1/3} a_{ij}$ with $a_{ij} = \sqrt{a_i a_j}$.
The time interval between two conjunctions (i.e., the synodic period)  is approximately given by
\begin{equation}
    T_\text{syn} \simeq \frac{2\pi}{(\dv*{\Omega_\text{K}}{r})b}
    \simeq \frac{4\pi a_{ij}}{3b\Omega_\text{K}}.
\end{equation}
Therefore, the expansion rate of the separation is found to be
\begin{equation}
     \dv{b}{t} \sim \frac{\Delta b}{T_\text{syn}}
           = 7 \qty( \frac{b}{r_\text{H}} )^{-4}
	     \qty( \frac{r_\text{H}}{a_{ij}} )^2 v_\text{K}.
\end{equation}
When the converging speed of the two orbits $\Delta v_\text{mig}$ 
($\equiv |(\dot{a}_j)_\text{mig}-(\dot{a}_i)_\text{mig}|$) becomes equal to $\dv*{b}{t}$, the two embryos are assumed to stop approaching, ending up captured in a mean-motion resonance with each other.
The resultant separation is, thus,
\begin{equation}
    b_\text{trap} = 0.16  \qty( \frac{M_i+M_j}{M_\oplus} )^{1/6}
                       \qty( \frac{\Delta v_\text{mig}}{v_\text{K}} )^{-1/4} r_\text{H}.
\end{equation}
Note that we do not specify in which mean motion resonance those two embryos are really captured, but assume the mean motion resonance with the width nearly equal to $b_\text{trap}$.
In the planetary population synthesis calculations, we calculate $b_\text{trap}$ for all the converging pairs. If $b < b_\text{trap}$ is satisfied, the pair is treated as captured in the resonance.
If $b_\text{trap} < 2\sqrt{3}r_\text{H}$, the two embryos or planets collide and merge with each other.

After trapped in the resonance, the pair migrates keeping the ratio of their orbital periods (i.e., the ratio of their semi-major axes) unchanged.
In that case, their migration rates are different from Eq.~\eqref{eq:da_dt}.
We calculate the loss of the angular momenta of the resonantly trapped planets via the interaction between the respective planets and the disc, 
and, then, redistribute the loss to both planets so that the planets migrate with the fixed semi-major axis ratio.
The migration rate of the resonantly trapped planet $i$
denoted by $v_\text{mig,trap}$ and that of the planet $j$ 
by $C_{ij}v_\text{mig,trap}$ ($C_{ij} \equiv a_j/a_i$),
that angular momentum loss rate, $\dot{\mathcal{L}}_\text{trap}$, can be expressed as
\begin{align}
    \dot{\mathcal{L}}_\text{trap} 
    &= \dv{t}(M_i \sqrt{GM_*a_i}+M_j\sqrt{GM_*a_j}) \\
    &= \frac{M_i \sqrt{GM_*a_j}+C_{ij}M_j\sqrt{GM_*a_i}}{2a_{ij}} v_\text{mig,trap}.
\end{align}
Here we assume that the eccentricities of both planets are negligible.
Thus, $v_\text{mig,trap}$ becomes
\begin{equation}
    v_\text{mig,trap} = 
    \frac{2a_{ij} (\Gamma_{i}+\Gamma_{j})}{M_i \sqrt{GM_*a_j}+C_{ij}M_j\sqrt{GM_*a_i}},
\end{equation}
where $\Gamma_i$ and $\Gamma_j$ are the torques exerted on the planets $i$ and $j$, respectively, by disc gas, calculated with Eq.~\eqref{eq:Gamma_typeII}.

The orbital migration terminates at the disc inner edge. 
The subsequent planets moving inward are trapped in the resonance, and multiple planets line up near the disc inner edge.
However, if a heavy enough planet joins such a resonance chain,
the planet pushes planets ahead of itself and, in particular, the innermost one into the disc cavity.
Following \cite{Ida+2013}, we adopt the condition for this `leakage' of the planet based on \cite{Ogihara+2010}; namely, once the following condition is satisfied, the planet at the disc edge halts the migration of the subsequent planet:
\begin{equation}
    \frac{e_1}{2\pi}\frac{M_1 \mathcal{L}_1}{\tau_e(M_1)}
    -\sum_{i=1}^{N_\text{edge}} \frac{M_i \mathcal{L}_i}{2\tau_\text{mig}(M_i)} > 0,
    \label{eq:ecc_trap_cond}
\end{equation}
where $M_i$ is the mass of the $i$-th planet ($i=1$ is the planet at the inner edge), $\mathcal{L}_i \sim \sqrt{GM_*a_i}$ is the orbital angular momentum, $\tau_\text{mig}$ is the migration timescale, $\tau_e$ is the eccentricity damping timescale due to migration, which is given by~\cite{Tanaka+Ward2004} as
\begin{equation}
    \tau_e = \frac{1}{0.78}\qty(\frac{M_\text{p}}{M_*})^{-1}
    \qty(\frac{\Sigma_\text{g} a^2}{M_*})^{-1}
    \qty(\frac{h}{a})^4 \Omega_\text{K}^{-1},
\end{equation}
$e_1$ is the eccentricity of the inner-most planet, which is fitted by ref.~\cite{Ogihara+2010} as
\begin{equation}
    e_1 = 0.02 \qty(\frac{\tau_e/\tau_\text{mig}}{10^{-3}})^{1/2},
\end{equation}
and, finally, $N_\text{edge}$ is the number of planets trapped near the disc inner edge.
If Eq.~\eqref{eq:ecc_trap_cond} is not satisfied, the inner-most planet is pushed into the disc cavity.
We assume that the semi-major axis ratio for this pushed-out planet and the next planet is kept constant, until orbital crossing or a giant impact occurs.

\subsubsection*{Case including giant planets}
The resonance trapping conditions for a pair of an embryo and a giant planet and that of two giant planets are the same as in the two-embryo case.
The difference is in the outcome that happens when $b_\text{trap} < 2 \sqrt{3} r_\text{H}$. Note that these cases are rare in planetary synthesis models for M dwarfs because giant planets are rarely formed.

A pair of two giant planets, if $b_\text{trap} < 2\sqrt{3}r_\text{H}$, undergo close encounters.
The post-process is calculated following the procedure for dynamical interactions between protoplanets (see Section~\ref{sec:dyanamic_interact} (ii)).

For a pair of an embryo and a giant planet, the trapping condition is calculated based on \cite{Shiraishi+Ida2008}, which gives a criterion for an embryo to enter the feeding zone of a giant planet.
This condition is expressed by comparing the decreasing rate in the separation of two bodies relative to the Hill radius of the giant planet $v_\text{H}$ and the eccentricity damping rate of the embryo $v_\text{damp}$.
Given that an embryo of mass $M_\text{E}$ and semi-major axis $a_\text{E}$ interacts with a giant planet of $M_\text{G}$ and $a_\text{G}$,
$v_\text{H}$ is given by 
\begin{align}
    v_\text{H} = \dv{\tilde{b}^2}{t}
    = \frac{2\tilde{b}}{h_\text{G}}\frac{a_\text{E}}{a_\text{G}}
    \qty(
    \frac{\dot{a}_\text{E}}{a_\text{E}} - 
    \frac{\dot{a}_\text{G}}{a_\text{G}}
    )
    -\frac{2\tilde{b}^2}{3}\frac{\dot{M}_\text{G}}{M_\text{G}},
\end{align}
where $\tilde{b} = (a_\text{E}-a_\text{G})/h_\text{G}a_\text{G}$ with $h_\text{G}=(M_\text{G}/3M_*)^{1/3}$.
On the other hand, $v_\text{damp}$ is expressed as 
\begin{equation}
    v_\text{damp} = -\frac{(e_\text{E}/h_\text{G})}{\tau_\text{damp}}
\end{equation}
with the damping timescale given by~\citep{Artymowicz1993,Iwasaki+2002}
\begin{align}
\begin{split}
 \tau_\text{damp} &= 400~\si{yr} 
 \qty( \frac{H_\text{disc}/a}{0.03} )^{4}
 \qty( \frac{\Sigma_\text{g}}{1000~\si{g/cm^2}} )^{-1} \\
 & \quad \times \qty( \frac{M_\text{E}}{M_\oplus} )^{-1} 
\qty( \frac{a_\text{E}}{1~\si{au}} )^{-1/2}
		\qty( \frac{M_*}{M_\odot} )^{-1/2}.
\end{split}		
\end{align}
If $|v_\text{H}|>|v_\text{damp}|$ is satisfied, the embryo enters the feeding zone of the giant planet and undergoes close encounters. 
The post-process is also calculated following the procedure given in Section~\ref{sec:dyanamic_interact} (ii).

\subsection*{Dynamic interaction of multi-body systems and its outcome}\label{sec:dyanamic_interact}
As the disc gas depletes, the typical timescale for eccentricity damping via gas drag becomes larger than the timescale for the orbit destabilisation and crossing of the closest pair to happen (called the orbital crossing timescale).
In this case, orbital crossing occurs between the pair.
We use the semi-analytical model from \cite{Ida+Lin2010} and \cite{Ida+2013} regarding the orbital repulsion, merging events, and gravitational scatterings to calculate the resultant semi-major axis and planetary mass \citep[see][for the details]{Ida+Lin2010,Ida+2013}.
Here we briefly summarise the numerical procedure.

The treatment of dynamical interactions differs depending on the number of giant planets in the system $N_\text{giant}$: (i) $N_\text{giant}=0,1$, (ii) $N_\text{giant}=2$, and (iii) $N_\text{giant} \ge 3$.
Here we define `giant planets' as the planets that satisfy both conditions (1) $M_\text{p}>30M_\oplus$ and (2) $e_\text{esc} > 1$, where
\begin{align}
    e_\text{esc} &= \frac{v_\text{esc}}{v_\text{K}}
    = \frac{\sqrt{2GM_\text{p}/R_\text{p}}}{\sqrt{GM_*/a_\text{p}}} \notag \\
\begin{split}
    &=  1.6 \qty( \frac{M_\text{p}}{M_\text{J}} )^{1/3}
            \qty( \frac{\bar{\rho}}{1~\si{g/cm^3}} )^{1/6}\\
    & \qquad \qquad 
    \times \qty( \frac{a_\text{p}}{1~\si{au}} )^{1/2}
    \qty( \frac{M_*}{M_\odot})^{-1/2},
\end{split}            
\end{align}
with $M_\text{J}$ the Jovian mass and $\bar{\rho}$ the mean density of the planet.

\begin{enumerate}[label=(\roman*)]
\item If the system has no or only one giant planet, we first calculate the orbital crossing timescale for every adjacent pair of embryos $(i,j)$.
The timescale $\tau_\text{cross}$ follows the fitting formula given by \cite{Zhou+2007} as
\begin{equation}
    \log(\frac{\tau_\text{cross}}{T_\text{K}}) = A + B \log(\frac{b}{2.3r_\text{H}}),
    \label{eq:tau_cross}
\end{equation}
where $T_\text{K}$ is the Keplerian period at the semi-major axis of $a=\sqrt{a_j a_i}$, $b=|a_i-a_j|$, $r_\text{H} = ((M_i + M_j)/3M_*)^{1/3} \min(a_i,a_j)$, and
\begin{align}
\begin{split}
 A &= -2.0 + e_0 -0.27 \log \mu \\
 &\quad + 0.51i_0+0.19i_0\log \mu + 0.03i_0 (\log \mu)^2 ,
\end{split} 
\\
\begin{split}
 B &= 18.7 + 1.1\log\mu - (16.8 + 1.2\log \mu) e_0 \\
 & \quad - 0.28i_0 + 0.19i_0 \log\mu,
\end{split} 
\\
 e_0 &= \frac{1}{2}\frac{e_i + e_j}{b}a, \quad
 \mu  = \frac{1}{2} \frac{M_i + M_j}{M_*}. 
\end{align}
Here $i_0$ is the mean inclination of the two planets in a unit of degree. The inclination of a planet in a unit of radian is set to the half of the eccentricity. The planetary eccentricity before any orbital crossing events is assumed to follow the Rayleigh distribution with root mean square (rms) of $\sigma = (M_i/3M_*)^{1/3}$.
Then, after the time interval equal to $\tau_\text{cross}$, the pair $(i,j)$ undergoes orbital crossing or sometimes ends up merging.
Their resultant masses, semi-major axes, and eccentricities are calculated following the procedure presented in \cite{Ida+Lin2010}.
They are evaluated so that each of the total mass, orbital energy, and Laplace-Runge-Lenz vector is conserved.
Finally, if the orbit of the embryo is within $3.5R_\text{H}$ of the giant planet, the embryo is scattered by the giant planet, and its semi-major axis and eccentricity are modified again, following ref.\cite{Ida+2013}.

\item If the system has two giant planets (1 and 2) and their separation is $b = |a_1-a_2| < 2\sqrt{3}r_\text{H}$, with $r_\text{H}=((M_1+M_2)/3M_*)^{1/3}\sqrt{a_1 a_2}$, then, 
the orbital instability occurs.
The orbital elements after the instability are calculated following `Two Giants Case' in \cite{Ida+2013}.
If the orbital instability occurs, all other embryos are assumed to be ejected from the system.
Otherwise, the interactions between embryos are calculated in the same way as in Case (i).

\item If the system has more than two giant planets, the orbital crossing timescale is calculated for every pair of giant planets using Eq.~\eqref{eq:tau_cross}.
If any of the derived $\tau_\text{cross}$ is larger than the total integration time, the giant planets do not interact with each other, and only the interaction between embryos are calculated following Case (i).
Otherwise, the orbital instability occurs after $\tau_\text{cross}$, and the resultant masses and orbital elements are derived from the `Three Giants Case' model in \cite{Ida+2013}.
All the other embryos are ejected from the system in this case.
\end{enumerate}

\subsection*{Initial Conditions and Parameters} \label{sec:setting}
To start the planetary population synthesis simulations, we perform random samplings of the initial mass $M_\text{disc}$, radius $r_\text{disc}$, metallicity [Fe/H], and inner edge radius $r_\text{in}$ of the protoplanetary gas disc, 
the external photo-evaporation rate $\dot{M}_\text{wind}$, 
and the initial masses and semi-major axes of embryos in the following way.
We use the default subroutine \verb|random_number| in FORTRAN90 to generate random numbers of uniform distribution in $[0,1]$.
To generate random numbers following the normal and Rayleigh distributions, we use Monty Python method and inverse transform method, respectively.

\subsubsection*{Initial conditions for protoplanetary disc} \label{sec:init_cond_disc}
We determine the initial properties of the protoplanetary disc by scaling recent observation results for stars of $\sim$~1~$M_\odot$. We adopt the fitting formula for the disc gas mass for $1M_\odot$ stars,
the log-normal distribution with the mean $\log(\mu/M_*)=-1.49$ and the standard deviation $\sigma = 0.35$, which \citet{Emsenhuber+2021b} derived from observational results of \cite{Tychoniec+2018}. 
The minimum and maximum disc masses are set to $\num{4e-3}M_\odot$ and $0.16M_\odot$, respectively, which roughly correspond to the lightest and heaviest samples in \cite{Tychoniec+2018}.
We also assume that the mean, minimum, and maximum disc masses are proportional to the stellar masses~\citep{Andrews+2013}.
The disc gas radius is calculated with
\begin{equation}
    r_\text{disc} = 10 \qty(\frac{M_\text{disc}}{\num{2e-3}M_\odot})^{0.625}~\si{au},
\end{equation}
which is taken from the observational trend derived in \cite{Andrews+2010}.

The disc metallicity [Fe/H] follows the normal distribution with $\mu=-0.02$ and $\sigma = 0.22$ derived from \cite{Santos+2005}.
The range of the value is limited to $-0.6<$[Fe/H]$<0.5$.
We use the same distribution regardless of the stellar mass.

We assume that the disc inner edge 
locates at the corotation radius where the Kepler rotation period is equal to the rotation period of the central star.
Here the stellar rotation period is assumed to follow the log-normal distribution with $\log (\mu \si{[days]})=0.676$ and $\sigma = 0.306$ based on the observational results of young stellar objects by \cite{Venuti+2017}.
The minimum of $r_\text{in}$ is set to the initial stellar radius $R_*$.
We also use the same distribution for all stellar types.

The external photo-evaporation rate $\dot{M}_\text{wind}$ generally depends on the population of nearby massive stars.
Following \cite{Burn+2021}, we set the distribution of $\dot{M}_\text{wind}$ so that the mean value of the resultant disc lifetime locates at $\sim$3Myr~\citep{Mamajek2009,Ansdell+2018} and that it has a deviation of about half an order, regardless of the stellar mass.
Since the disc gas radius $r_\text{disc}$ is determined only by the disc mass $M_\text{disc}$, the disc lifetime depends on $M_\text{disc}$ and $\dot{M}_\text{wind}$ for given $\alpha_\text{acc}$.
Then we find that, for a star of $0.3M_\odot$ and $\alpha_\text{acc}=\num{2e-3}$, the log-normal distribution with $\log(\mu ~\si{[M_\odot/yr]})=-6.0$ and $\sigma = 0.5$ accounts for the above distribution.
Also, the stellar mass dependence of $\dot{M}_\text{wind} \propto M_*^{1.4}$ is found to be suitable for the star in the range of $0.1M_\odot \le M_* \le 0.5M_\odot$.

\subsubsection*{Initial conditions for planetary embryos} \label{sec:init_cond_planet}

Initially, 50 planetary embryos with mass of 0.01~$M_\oplus$ are placed log-uniformly from $r_\text{in}$ to $r_\text{solid}$.
Here the initial separations of all the adjacent embryos are larger than the feeding zone width ($10r_\text{H}$) for the local isolation mass $M_\text{iso}$ given by~\citep{Kokubo+Ida2002}
\begin{equation}
    M_\text{iso} = 0.16 \qty(\frac{\Sigma_\text{s}}{10~\si{g.cm^{-2}}})^{3/2}
    \qty(\frac{a}{1~\si{au}})^{3/4} \qty(\frac{M_*}{M_\odot})^{1/2} ~M_\oplus.
\end{equation}
Therefore, in quite massive solid discs, the initial number of planetary embryos can be smaller than 50.

\subsubsection*{Input parameters}
The parameters and their fiducial values are summarised in Table~\ref{tab:parameter}.
These values are used in our calculations unless otherwise mentioned.

\subsection*{Probability density}
To show the distribution of water mass fraction in the synthesised planets (Figs.~\ref{fig:hist_M03} and \ref{fig:hist_Mstar}), we use the probability densities (PDs) calculated by
\begin{equation}
    {\rm PD}(i)
    = \frac{N_i}{N_{\rm tot}\Delta \log (M_{\rm water}/M_{\rm core})},
\end{equation}
where $N_i$ is the number of planets in the $i$-th bin, $N_{\rm tot}=\sum{N_i}$ is the total number of planets, and $\Delta \log(M_{\rm water}/M_{\rm core})$ is the bin width.

\section*{Data availability}

All data from the simulation are available at \url{https://github.com/TadahiroKimura/Kimura-Ikoma2022}.
Soure Data for each figure is provided with this paper.


\section*{Code availability}
The numerical code used in the current study is available from the corresponding author upon request only for the purpose of reproducing our results.

\section*{Acknowledgements}

This work is supported by JSPS KAKENHI Nos.~JP18H05439, JP21H01141 and JP22J11725. TK is a JSPS Research Fellow, and also
supported by International Graduate Program for Excellence in Earth-Space Science (IGPEES).

\section*{Author contributions}
Both authors contributed equally to this work. M.I. conceived the original idea  and supervised this project. T.K. developed the entire model of planetary population synthesis partly using a few modules that M.I. had developed. T.K. carried out the numerical simulations and analyzed the simulation results. Both authors discussed the results and implications and wrote the paper.

\section*{Corresponding author}
Correspondence and requests for materials should be addressed to T. Kimura.

\section*{Competing interests}
The authors declare no competing interests.

\begin{table*}
    \centering
    \caption{Parameters used in calculations}   
    \label{tab:parameter}
    \begin{tabular}{c|>{\centering}p{5cm}|c|c}\hline
         Symbol & Meaning & Value & Eq. mainly used \\
         \hline
         $\alpha_\text{acc}$ & 
         parameter for effective turbulent viscosity & $\num{2.0e-3}$ &
         Eq.~\eqref{eq:dSigma_dt_basic}\\
         $\alpha_\text{vis}$ & 
         parameter for turbulent viscosity &
         $\num{2.0e-4}$ & Eq.~\eqref{eq:Kanagawa}\\ 
         $N_\text{grid,disc}$ & 
         number of grids for gas disc &
         500 & \\
         $N_\text{grid,solid}$ & 
         number of grids for solid disc &
         1000 & \\
         $r_\text{max}$ &
         outer boundary radius for gas disc &
         1000~au & Eq.~\eqref{eq:dSigma_pe_ext} \\
         $r_\text{solid}$ &
         solid disc radius &
         $0.5r_\text{disc}$ & Eq.~\eqref{eq:Sigma_solid} \\
         $\Phi$ &
         ionising EUV photon luminosity &
         $\num{1.0e40}~\si{s^{-1}}$ & Eq.~\eqref{eq:n0} \\
         $m_\text{plt}$ & 
         planetesimal mass &
         $\num{1.0e20}$~g & Eq.~\eqref{eq:tilde_e_plt} \\
         $\rho_\text{plt}$ & 
         planetesimal material density &
         $3.0~\si{g.cm^{-3}}$ & Eq.~\eqref{eq:tilde_e_plt} \\
        $C_\text{rock}$ & 
        specific heat of solid core for constant volume& 
        $\SI{1.2e7}{erg/(g.K)}$ & Eq.~\eqref{eq:Lcool} \\ 
         $K_t$ & 
         typical value of $K$ for which the corotation torque becomes ineffective & 
         20.0 & Eq.~\eqref{eq:Gamma_typeII} \\
         %

         \hline
    \end{tabular}
\end{table*}

\clearpage
\section*{References}
%


\end{document}